\documentclass[prd,aps,showpacs,epsf,floats,onecolumn]{revtex4}%
\usepackage{amssymb}
\usepackage{amsfonts}
\usepackage{amsmath}
\usepackage{graphicx}%
\providecommand{\U}[1]{\protect\rule{.1in}{.1in}}
\begin{document}
\title{\textbf{Geometric aspects of mixed quantum states inside the Bloch sphere}}
\author{\textbf{Paul M. Alsing}$^{1}$, \textbf{Carlo Cafaro}$^{2,3}$, \textbf{Domenico
Felice}$^{3,4}$, \textbf{Orlando Luongo}$^{3,5,6,7,8}$}
\affiliation{$^{1}$Air Force Research Laboratory, Information Directorate, 13441 Rome, New
York, USA}
\affiliation{$^{2}$University at Albany-SUNY, Albany, NY 12222, USA}
\affiliation{$^{3}$SUNY Polytechnic Institute, Utica, NY 13502, USA}
\affiliation{$^{4}$Scuola Militare Nunziatella, Napoli 80132, Italy}
\affiliation{$^{5}$Universit\`{a} di Camerino, Camerino 62032, Italy}
\affiliation{$^{6}$INAF-Osservatorio Astronomico di Brera, Milano 20121, Italy}
\affiliation{$^{7}$INFN,\ Sezione di Perugia, Perugia 06123 , Italy}
\affiliation{$^{8}$Al-Farabi Kazakh National University, Almaty 050040, Kazakhstan}

\begin{abstract}
When studying the geometry of quantum states, it is acknowledged that mixed
states can be distinguished by infinitely many metrics. Unfortunately, this
freedom causes metric-dependent interpretations of physically significant
geometric quantities such as complexity and volume of quantum states. In this
paper, we present an insightful discussion on the differences between the
Bures and the Sj\"{o}qvist metrics inside a Bloch sphere. First, we begin with
a formal comparative analysis between the two metrics by critically discussing
three alternative interpretations for each metric. Second, we illustrate
explicitly the distinct behaviors of the geodesic paths on each one of the two
metric manifolds. Third, we compare the finite distances between an initial
and final mixed state when calculated with the two metrics. Interestingly, in
analogy to what happens when studying topological aspects of real Euclidean
spaces equipped with distinct metric functions (for instance, the usual
Euclidean metric and the taxicab metric), we observe that the relative ranking
based on the concept of finite distance among mixed quantum states is not
preserved when comparing distances determined with the Bures and the
Sj\"{o}qvist metrics. Finally, we conclude with a brief discussion on the
consequences of this violation of a metric-based relative ranking on the
concept of complexity and volume of mixed quantum states.

\end{abstract}

\pacs{Quantum Computation (03.67.Lx), Quantum Information (03.67.Ac), Riemannian
Geometry (02.40.Ky).}
\maketitle

\bigskip\pagebreak

\section{Introduction}

It is established that there exist infinitely many distinguishability metrics
for mixed quantum states \cite{karol06}. For this reason, there is a certain
degree of arbitrariness in selecting the metric when characterizing physical
aspects of quantum states in mixed states. In particular, this freedom can
cause metric-dependent explanations of geometric quantities with a clear
physical significance, including complexity
\cite{cafaroPRD22,chapman18,ruan21,gu12,iaconis21,brandao21,bala22,belin22,omidi23}
and volume
\cite{karol98,karol99,felice17,felice18,kz01,kz03,kz03A,andai04,ye09,ye10,singh14,siu20}
of quantum states. Two examples of metrics for mixed quantum states are the
Bures \cite{bures69,uhlmann76,hubner92,sam94} and the Sj\"{o}qvist
\cite{erik20} metrics. In Ref. \cite{cafaro23EPJ}, we proposed a first
explicit characterization of the Bures and Sj\"{o}qvist metrics over the
manifolds of thermal states for specific spin qubit and superconducting flux
qubit Hamiltonian models. We observed that while both metrics become the
Fubini-Study metric in the asymptotic limiting case of the inverse temperature
approaching infinity for both Hamiltonian models, the two metrics are
generally distinct when far from the zero-temperature limit. The two metrics
differ in the presence of a nonclassical behavior specified by the
noncommutativity of neighboring mixed quantum states. \ Such a
noncommutativity, in turn, is taken into account by the two metrics
differently. As a follow up of our work in \cite{cafaro23EPJ}, we used the
concept of decompositions of density operators by means of ensembles of pure
quantum states to present in Ref. \cite{alsing23} an unabridged mathematical
investigation on the relation between the Sj\"{o}qvist metric and the Bures
metric for arbitrary nondegenerate mixed quantum states. Furthermore, to
deepen our comprehension of the difference between these two metrics from a
physics standpoint, we compared the general expressions of these two metrics
for arbitrary thermal quantum states for quantum systems in equilibrium with a
reservoir at non-zero temperature. Then, for clarity, we studied \ the
difference between these two metrics in the case of a spin-qubit in an
arbitrarily oriented uniform and stationary external magnetic field in thermal
equilibrium with a finite-temperature bath. Finally, we showed in Ref
\cite{alsing23} that the Sj\"{o}qvist metric does not satisfy the so-called
monotonicity property \cite{karol06}, unlike the Bures metric. An interesting
observable consequence, in terms of complexity behaviors, of this freedom in
choosing between the Bures and Sj\"{o}qvist metrics was reported in Ref.
\cite{cafaroPRD22}. There, devoting our attention to geodesic lengths and
curvature properties for manifolds of mixed quantum states, we recorded a
softening of the information geometric complexity \cite{cafaro07,cafarothesis}
on the Bures manifold compared to the Sj\"{o}qvist manifold.

In this paper, motivated by our findings in Refs.
\cite{cafaroPRD22,cafaro23EPJ,alsing23}, we present a more in depth conceptual
discussion on the differences between the Bures and the Sj\"{o}qvist metrics
inside a Bloch sphere. To achieve this goal, we first begin by presenting in
Section II a formal comparative analysis between the two metrics. This
analysis is based upon a critical discussion on three different alternative
interpretations for each one of the two metrics. We then continue in Section
III with an explicit illustration of the different behaviors of the geodesic
paths on each one of the two metric manifolds. In the same section, we also
compare the finite distances between an initial and final mixed state when
calculated by means of the two metrics. Inspired by what happens when studying
topological aspects of real Euclidean spaces equipped with distinct metric
functions (for instance, the usual Euclidean metric and the taxicab metric),
we observe in Section IV that the relative ranking based on the concept of
finite distance among mixed quantum states is not preserved when comparing
distances determined with the Bures and the Sj\"{o}qvist metrics. We then
discuss in Section IV the consequences of this violation of a metric-based
relative ranking on the concept of complexity and volume of mixed quantum
states, along with other geometric peculiarities of the Bures and the
Sj\"{o}qvist metrics inside a Bloch sphere. Our concluding remarks appear in
Section V. Finally, for the ease of presentation, some more technical details
appear in Appendix A and B.

Before transitioning to our next section, we acknowledge that the presentation
of the content of this paper is more suitable for specialists interested in
the geometric aspects of mixed quantum states. However, for interested readers
who are not so familiar with the topic, we suggest Ref. \cite{karol06} for a
general introduction to the geometry of quantum states. Furthermore, for a
tutorial on the geometry of Bures and Sj\"{o}qvist manifolds of mixed states,
we refer to Ref. \cite{cafaro23EPJ}. Finally, for a partial list of more
technical applications of the Bures and Sj\"{o}qvist metrics in quantum
information science, we suggest Refs.
\cite{hubner93,slater98,dittmann99,zanardi07A,zanardi07B,safranek17,hornedal22,huang23}
and Refs. \cite{silva21,dasilva21,kim21,mera22,daniel23,chien23}, respectively.

\begin{table}[t]
\centering
\par%
\begin{tabular}
[c]{c|c|c}\hline\hline
Metric & Riemannian property & Monotonicity\\\hline
Bures & Yes & Yes\\
Hilbert-Schmidt & Yes & No\\
Sj\"{o}qvist & Yes & No\\
Trace & No & Yes\\\hline
\end{tabular}
\caption{Examples of metrics in the space of quantum states characterized in
terms of Riemannian property and monotonicity.}%
\end{table}

\section{Line elements}

In this section, we begin with a presentation of a formal comparative analysis
between the Bures and the Sj\"{o}qvist metrics inside a Bloch sphere. For
completeness, we first mention in Table I some examples of metrics for mixed
quantum states and characterize them in terms of their Riemannian and
monotonicity properties. For more details on the notion of monotonicity and
Riemannian property for quantum metrics, we refer to Ref. \cite{karol06}.
Returning to our main analysis, we focus here on the geometry of single-qubit
mixed quantum states characterized by density operators on a two-dimensional
Hilbert space. In this case, an arbitrary density operator $\rho$ can be
written as a decomposition of four-linear operators (i.e., four $(2\times
2)$-matrices) given by the identity operator \textrm{I} and the usual Pauli
vector operator $\vec{\sigma}\overset{\text{def}}{=}\left(  \sigma_{x}\text{,
}\sigma_{y}\text{, }\sigma_{z}\right)  $ \cite{nielsen00}. Explicitly, we have%
\begin{equation}
\rho\overset{\text{def}}{=}\frac{1}{2}\left(  \mathrm{I}+\vec{p}\cdot
\vec{\sigma}\right)  \text{,} \label{one}%
\end{equation}
where $\vec{p}\overset{\text{def}}{=}p\hat{p}$ denotes the three-dimensional
Bloch vector. Note that $p$ is the length $\left\Vert \vec{p}\right\Vert
\overset{\text{def}}{=}\sqrt{\vec{p}\cdot\vec{p}}$ of the polarization vector
$\vec{p}$, while $\hat{p}$ is the unit vector. Following the vectors and
one-forms notation along with the line of reasoning presented in Refs.
\cite{sam95A,sam95B,cafaro12}, we can formally recast $\vec{p}$ and
$\vec{\sigma}$ in\ Eq. (\ref{one}) as%
\begin{equation}
\vec{p}\overset{\text{def}}{=}\sum_{i=1}^{3}p^{i}\hat{e}_{i}\text{, and }%
\vec{\sigma}\overset{\text{def}}{=}\sum_{i=1}^{3}\sigma^{i}\hat{e}_{i}\text{,}%
\end{equation}
respectively. Observe that $\left\{  \hat{e}_{i}\right\}  _{1\leq i\leq3}$ is
a set of orthonormal three-dimensional vectors satisfying $\hat{e}_{i}%
\cdot\hat{e}_{j}=\delta_{ij}$, with $\delta_{ij}$ being the Kronecker delta
symbol. Moreover, we have $\left(  \sigma^{1}\text{, }\sigma^{2}\text{,
}\sigma^{3}\right)  \overset{\text{def}}{=}\left(  \sigma_{x}\text{, }%
\sigma_{y}\text{, }\sigma_{z}\right)  $. For pure states, $\rho=\rho^{2}$,
\textrm{tr}$\left(  \rho\right)  =\mathrm{tr}\left(  \rho^{2}\right)  =1$, and
$p=1$. Therefore, pure states are located on the surface of the unit
two-sphere. For mixed states states, instead, $\rho\neq\rho^{2}$,
\textrm{tr}$\left(  \rho\right)  =1$, and \textrm{tr}$\left(  \rho^{2}\right)
\leq1$. Since \textrm{tr}$\left(  \rho^{2}\right)  =p^{2}$, we have $p\leq1$.
Therefore, mixed quantum states are located inside the unit two-sphere, i.e.,
they belong to the interior of the Bloch sphere.

In next two subsections, we study the geometric aspects of the interior of the
Bloch sphere specified by the Bures and the Sj\"{o}qvist line elements, respectively.

\subsection{The Bures line element}

In the case of the Bures geometry, the infinitesimal line element
$ds_{\mathrm{Bures}}^{2}\left(  \vec{p}\text{, }\vec{p}+d\vec{p}\right)  $
between two neighboring mixed states $\rho$ and $\rho+d\rho$ corresponding to
Bloch vectors $\vec{p}$ and $\vec{p}+d\vec{p}$ is given by
\cite{hubner92,sam95A,sam95B}%
\begin{equation}
ds_{\mathrm{Bures}}^{2}\left(  \vec{p}\text{, }\vec{p}+d\vec{p}\right)
=\frac{1}{4}\left[  \frac{\left(  \vec{p}\cdot d\vec{p}\right)  ^{2}}{1-p^{2}%
}+d\vec{p}\cdot d\vec{p}\right]  =\frac{1}{4}\left[  \frac{dp^{2}}{1-p^{2}%
}+p^{2}(d\hat{p}\cdot d\hat{p})\right]  \text{.} \label{bures1}%
\end{equation}
The equality between the first and second expressions of $ds_{\mathrm{Bures}%
}^{2}\left(  \vec{p}\text{, }\vec{p}+d\vec{p}\right)  $ in Eq. (\ref{bures1})
can be checked by first noting that $\hat{p}\cdot\hat{p}=1$ implies $\hat
{p}\cdot d\hat{p}=0$. This, in turn, yields the relations $d\vec{p}\cdot
d\vec{p}=dp^{2}+p^{2}d\hat{p}\cdot d\hat{p}$ and $\left(  \vec{p}\cdot
d\vec{p}\right)  ^{2}=p^{2}dp^{2}$. Finally, the use of these two identities
allows us to arrive at the equality between the two expressions in Eq.
(\ref{bures1}).

In what follows, we propose three interpretations for the Bures metric which
originate from a critical reconsideration of the original work by Braunstein
and Caves in Refs. \cite{sam95A,sam95B}.

\subsubsection{First interpretation}

We wish to critically discuss the structure of $ds_{\mathrm{Bures}}^{2}\left(
\vec{p}\text{, }\vec{p}+d\vec{p}\right)  $ in Eq. (\ref{bures1}). To interpret
the term $d\hat{p}\cdot d\hat{p}$ in $ds_{\mathrm{Bures}}^{2}\left(  \vec
{p}\text{, }\vec{p}+d\vec{p}\right)  $, it is convenient to recast the
polarization vector in spherical coordinates as $\vec{p}=\left(  p^{1}\text{,
}p^{2}\text{, }p^{3}\right)  \overset{\text{def}}{=}\left(  p\sin\left(
\theta\right)  \cos\left(  \varphi\right)  \text{, }p\sin\left(
\theta\right)  \sin\left(  \varphi\right)  \text{, }p\cos\left(
\theta\right)  \right)  $. Note that in spherical coordinates, we also have
$\left(  \hat{e}_{1}\text{, }\hat{e}_{2}\text{, }\hat{e}_{3}\right)  =\left(
\hat{e}_{r}\text{, }\hat{e}_{\theta}\text{, }\hat{e}_{\varphi}\right)  $. It
then follows that $d\vec{p}\cdot d\vec{p}=dp^{2}+p^{2}\left(  d\theta^{2}%
+\sin^{2}\left(  \theta\right)  d\varphi^{2}\right)  $ and $\left(  \vec
{p}\cdot d\vec{p}\right)  ^{2}=p^{2}dp^{2}$. Finally, noting that the unit
polarization vector is $\hat{p}=\left(  \sin\left(  \theta\right)  \cos\left(
\varphi\right)  \text{, }\sin\left(  \theta\right)  \sin\left(  \varphi
\right)  \text{, }\cos\left(  \theta\right)  \right)  $, we get after some
algebra that $d\hat{p}\cdot d\hat{p}=d\theta^{2}+\sin^{2}\left(
\theta\right)  d\varphi^{2}$. From this last relation, it clearly follows that
$d\hat{p}\cdot d\hat{p}$ represents the usual line element $d\Omega
^{2}\overset{\text{def}}{=}$ $d\theta^{2}+\sin^{2}\left(  \theta\right)
d\varphi^{2}$ on a unit two-sphere. Therefore, when using spherical
coordinates, the line element in Eq. (\ref{bures1}) can be recast as%
\begin{equation}
ds_{\mathrm{Bures}}^{2}\left(  \vec{p}\text{, }\vec{p}+d\vec{p}\right)
=\frac{1}{4}\left[  \frac{dp^{2}}{1-p^{2}}+p^{2}(d\theta^{2}+\sin^{2}\left(
\theta\right)  d\varphi^{2})\right]  \text{.} \label{burespherical}%
\end{equation}
From Eq. (\ref{bures1}), it happens that when $p$ is kept constant and equal
to $p_{0}$, a surface specified by the relation $p=p_{0}$ inside the Bloch
sphere exhibits the geometry of a two-sphere of area $4\pi p_{0}^{2}$. As
mentioned in Refs. \cite{sam95A,sam95B}, the term $dp^{2}/\left(
1-p^{2}\right)  $ in $ds_{\mathrm{Bures}}^{2}\left(  \vec{p}\text{, }\vec
{p}+d\vec{p}\right)  $ implies that the inside of the Bloch sphere is not
flat, but curved. Indeed, moving away from the origin of the sphere, the
circumference $C(p)\overset{\text{def}}{=}2\pi p$ of a circle of radius $p$ on
the two-sphere grows as $dC/ds\sim p^{\prime}\overset{\text{def}}{=}dp/ds$
with $s$ being the affine parameter. The distance $l\left(  p\right)
\overset{\text{def}}{=}\int_{s(0)}^{s(p)}$ $p^{\prime}/\left(  1-p^{2}\right)
^{1/2}ds$ from the center, instead, grows as $dl/ds\sim$ $p^{\prime}/\left(
1-p^{2}\right)  ^{1/2}\geq p^{\prime}$. Therefore, $l\left(  p\right)  $ grows
at a faster rate than $C(p)$. This discrimination in growth rates for $C(p)$
and $l(p)$ signifies that the interior of the Bloch sphere is curved.

\subsubsection{Second interpretation}

A second useful coordinate system to further gain insights into the Bures line
element in Eq. (\ref{bures1}) is specified by considering a change
of\textbf{\ }variables defined by $p\overset{\text{def}}{=}\sin\left(
\chi\right)  $ with $\chi$ being the hyperspherical angle with $0\leq\chi
\leq\pi/2$. \ In this set of coordinates, $4ds_{\mathrm{Bures}}^{2}\left(
\vec{p}\text{, }\vec{p}+d\vec{p}\right)  $ reduces to
\begin{equation}
4ds_{\mathrm{Bures}}^{2}=d\chi^{2}+\sin^{2}\left(  \chi\right)  d\hat{p}\cdot
d\hat{p}\text{,} \label{corre0}%
\end{equation}
with $d\hat{p}\cdot d\hat{p}=d\Omega^{2}\overset{\text{def}}{=}$ $d\theta
^{2}+\sin^{2}\left(  \theta\right)  d\varphi^{2}$. Note that Eq.
(\ref{corre0}) for the Bures metric is exactly the (intrinsic) metric on the
unit $3$-sphere $S^{3}$, where $\chi$, $\theta$, and $\varphi$ are the angular
coordinates on the sphere. For completeness, we remark that the metric for
the\textbf{ }$(N+1)$\textbf{-}sphere can be written in terms of the metric for
the\textbf{ }$N$-sphere, with the introduction of a new hyperspherical angle
\cite{karol06}. Two additional considerations are in order here. First, the
four-dimensional vector $x^{\mu}\left(  s\right)  \equiv\left(  x^{0}\left(
s\right)  \text{, }\vec{x}\left(  s\right)  \right)  $ with $0\leq\mu\leq3$
such that $dx^{\mu}dx_{\mu}=4ds_{\mathrm{Bures}}^{2}=d\chi^{2}+\sin^{2}\left(
\chi\right)  d\hat{p}\cdot d\hat{p}$ can be written as $x^{\mu}\left(
s\right)  =x^{0}\left(  s\right)  \hat{e}_{0}(s)+\vec{x}\left(  s\right)  $.
The quantity $x^{0}\left(  s\right)  \overset{\text{def}}{=}\chi\left(
s\right)  $ is the component of $x^{\mu}\left(  s\right)  $ along the
direction $\hat{e}_{0}(s)\overset{\text{def}}{=}\vec{\chi}/\chi$, with
$\hat{e}_{0}(s)\cdot\hat{e}_{i}(s)=0$ for any $1\leq i\leq3$. The
three-dimensional vector $\vec{x}\left(  s\right)  $ given by%
\begin{equation}
\vec{x}\left(  s\right)  =\int\sin\left[  \chi\left(  s\right)  \right]
\frac{d\hat{p}\left(  s\right)  }{ds}ds\text{,} \label{corre}%
\end{equation}
specifies the remaining three coordinates of $x^{\mu}\left(  s\right)  $ along
the directions $\hat{e}_{1}\left(  s\right)  \overset{\text{def}}{=}%
e_{r}\left(  s\right)  $, $\hat{e}_{2}\left(  s\right)  \overset{\text{def}%
}{=}e_{\theta}\left(  s\right)  $, and $\hat{e}_{3}\left(  s\right)
\overset{\text{def}}{=}e_{\varphi}\left(  s\right)  $. From Eq. (\ref{corre})
we point out the presence of a correlational structure between the motion
along $\hat{e}_{0}(s)$ and the \textquotedblleft spatial\textquotedblright%
\ directions $\hat{e}_{i}(s)$ with $1\leq i\leq3$. This correlational
structure is a manifestation of the fact that for the Bures geometry, radial
and angular motions inside the Bloch sphere are correlated since the dynamical
geodesic equations are specified by a set of second order coupled nonlinear
differential equations when using a set of spherical coordinates
\cite{cafaroPRD22}. Second, remembering that line element in the usual
cylindrical coordinates $\left(  \rho\text{, }\varphi\text{, }z\right)  $ is
$ds_{\text{cylinder}}^{2}=dz^{2}+d\Omega_{\text{cylinder}}^{2}$ where
$d\Omega_{\text{cylinder}}^{2}\overset{\text{def}}{=}d\rho^{2}+\rho
^{2}d\varphi^{2}$, we observe that the structure of the Bures line element
rewritten as in Eq. (\ref{corre0}) is suggestive of the structure of a line
element in the standard cylindrical coordinates once we make the connection
between the pair $\left(  \chi\text{, }d\Omega\right)  $ with the pair
$\left(  \rho\text{, }d\Omega_{\text{cylinder}}\right)  $. Then, one can link
a cylinder with a variable radius in the case of the Bures geometry. In
particular, it is worth mentioning at this point that the non constant radius
in the Bures case is upper bounded by the constant value that defines the
radius in the Sj\"{o}qvist geometry (as we shall see in the next subsection).
These geometric insights emerging from this simple change of coordinates would
lead one to reasonably expect different lengths of geodesic paths in the two
geometries studied here. This will be discussed in more detail in the next
section, however.

\subsubsection{Third interpretation}

An alternative third set of coordinates for the Bures line element in Eq.
(\ref{bures1}) is given by the four coordinates $x^{\mu}$ with $0\leq\mu\leq3$
given by $x^{\mu}=\left(  x^{0}\text{, }\vec{x}\right)  \overset{\text{def}%
}{=}\left(  \sqrt{1-p^{2}}\text{, }\vec{p}\right)  $ with $\vec{p}=p\hat{p}$.
Indeed, from $x^{0}=\sqrt{1-p^{2}}$, we get $\left(  dx^{0}\right)
^{2}=\left(  \vec{p}\cdot d\vec{p}\right)  ^{2}/(1-p^{2})$. Therefore, when
employing this coordinate system, the inside of the Bloch sphere can be
described by a three-dimensional surface defined by the constraint relation
$\left(  x^{0}\right)  ^{2}+\vec{x}\cdot\vec{x}=1$. Moreover, the geometry of
the surface is induced by the four-dimensional flat Euclidean line element
\cite{hubner92},%
\begin{equation}
4ds_{\mathrm{Bures}}^{2}=\left(  dx^{0}\right)  ^{2}+d\vec{x}\cdot d\vec
{x}\text{.} \label{hubi}%
\end{equation}
Notice that Eq. (\ref{hubi}) for the Bures metric is the (extrinsic) metric on
the unit $3$-sphere $S^{3}$ viewed as embedded in $%
\mathbb{R}
^{4}$. The geodesic paths emerging from $ds_{\mathrm{Bures}}^{2}$ in Eq.
(\ref{hubi}) are great circles on the $3$-sphere. In terms of the arc length
$s$, these geodesics can be recast as \cite{sam95A,sam95B}%
\begin{equation}
x^{\mu}\left(  s\right)  =u^{\mu}\cos(s)+v^{\mu}\sin(s)\text{,}
\label{geohubi}%
\end{equation}
where $u^{\mu}=\left(  u^{0}\text{, }\vec{u}\right)  \overset{\text{def}}%
{=}\left(  \cos\left(  \chi\right)  \text{, }\hat{n}\sin\left(  \chi\right)
\right)  $, $v^{\mu}=\left(  v^{0}\text{, }\vec{v}\right)  \overset
{\text{def}}{=}\left(  \sin\left(  \xi\right)  \text{, }\hat{m}\cos\left(
\xi\right)  \right)  $, $\hat{n}\cdot\hat{n}=\hat{m}\cdot\hat{m}=1$, and
$-\left(  \hat{n}\cdot\hat{m}\right)  \tan\left(  \chi\right)  =\tan\left(
\xi\right)  $. This last relation assures that $u^{\mu}\perp v^{\mu}$ so that
Eq. (\ref{hubi})\ is satisfied for $x^{\mu}\left(  s\right)  $ given in Eq.
(\ref{geohubi}).

We are now ready to critically discuss the Sj\"{o}qvist line element by
mimicking the discussion performed for the Bures line element.

\subsection{The Sj\"{o}qvist line element}

In the case of the Sj\"{o}qvist geometry, the infinitesimal line element
$ds_{\mathrm{Sj\ddot{o}qvist}}^{2}\left(  \vec{p}\text{, }\vec{p}+d\vec
{p}\right)  $ between two neighboring mixed states $\rho$ and $\rho+d\rho$
corresponding to Bloch vectors $\vec{p}$ and $\vec{p}+d\vec{p}$ is given by
\cite{erik20}%
\begin{equation}
ds_{\mathrm{Sj\ddot{o}qvist}}^{2}\left(  \vec{p}\text{, }\vec{p}+d\vec
{p}\right)  =\frac{1}{4}\left[  \frac{2p^{2}-1}{p^{4}\left(  1-p^{2}\right)
}\left(  \vec{p}\cdot d\vec{p}\right)  ^{2}+\frac{d\vec{p}\cdot d\vec{p}%
}{p^{2}}\right]  =\frac{1}{4}\left[  \frac{dp^{2}}{1-p^{2}}+(d\hat{p}\cdot
d\hat{p})\right]  \text{.} \label{erik1}%
\end{equation}
The equality between the first and second expressions of $ds_{\mathrm{Sj\ddot
{o}qvist}}^{2}\left(  \vec{p}\text{, }\vec{p}+d\vec{p}\right)  $ in Eq.
(\ref{erik1}) can be verified by first observing that $\hat{p}\cdot\hat{p}=1$
implies $\hat{p}\cdot d\hat{p}=0$. This, in turn, leads to the relations
$d\vec{p}\cdot d\vec{p}=dp^{2}+p^{2}d\hat{p}\cdot d\hat{p}$ and $\left(
\vec{p}\cdot d\vec{p}\right)  ^{2}=p^{2}dp^{2}$. Finally, exploiting these two
relations, we arrive at the equality between the two expressions in Eq.
(\ref{erik1}).

We remark that Sj\"{o}qvist in Ref. \cite{erik20} was the first to seek a
deeper understanding of the physics behind the metric, with the concept of
mixed state geometric phases playing a key role. However, for completeness, we
also point out that what we call \textquotedblleft Sj\"{o}qvist
interferometric metric\textquotedblright\ first appeared as a special case of
a more general family of metrics proposed in a more formal mathematical
setting in Refs. \cite{ole14a,ole14b} by Andersson and Heydari. In this
generalized setting, different metrics arise from different gauge theories,
they are specified by distinct notions of horizontality and, finally, they can
be well-defined for both nondegenerate and degenerate mixed quantum states.
Great part of the underlying gauge theory of this generalized family of
metrics was developed in Ref. \cite{ole15}. A suitable comprehensive reference
to read about such generalized family of metrics is Chapter 5 in Andersson's
thesis \cite{ole19} where, in particular, the singular properties of
Sj\"{o}qvist's metrics are discussed in Section 5.3.2. For further technical
details on this matter, we refer to Ref. \cite{ole19} and references therein.

\subsubsection{First interpretation}

We begin by noting that the term $dp^{2}/\left(  1-p^{2}\right)  $ in
$ds_{\mathrm{Sj\ddot{o}qvist}}^{2}\left(  \vec{p}\text{, }\vec{p}+d\vec
{p}\right)  $ implies that the inside of the Bloch sphere is not flat, but
curved. In particular, the interpretation of this term follows exactly the
discussion provided in the previous subsection for the Bures case. Moreover,
similarly to the Bures case, the term $d\hat{p}\cdot d\hat{p}$ remains the
standard line element $d\Omega^{2}\overset{\text{def}}{=}$ $d\theta^{2}%
+\sin^{2}\left(  \theta\right)  d\varphi^{2}$ on a unit two-sphere. Therefore,
when using spherical coordinates, the line element in Eq. (\ref{erik1}) can be
recast as%
\begin{equation}
ds_{\mathrm{Sj\ddot{o}qvist}}^{2}\left(  \vec{p}\text{, }\vec{p}+d\vec
{p}\right)  =\frac{1}{4}\left[  \frac{dp^{2}}{1-p^{2}}+d\theta^{2}+\sin
^{2}\left(  \theta\right)  d\varphi^{2})\right]  \text{.} \label{sspherical}%
\end{equation}
Unlike what happens in the Bures case, when $p$ is kept constant and equal to
$p_{0}$, a surface specified by the relation $p=p_{0}$ inside the Bloch sphere
exhibits the geometry of a two-sphere of area $4\pi$ in the Sj\"{o}qvist case.
The area $4\pi$ of this two-sphere is greater than the area $4\pi p_{0}^{2}$
that specifies the Bures case and, in addition, does not depend on the choice
of the constant value $p_{0}$ of $p$. This is a signature of the fact that, in
the Sj\"{o}qvist case, the accessible regions inside the Bloch sphere have
volumes greater than those specifying the Bures geometry. Indeed, this
observation was first pointed out in Ref. \cite{cafaroPRD22} and shall be
further discussed in the forthcoming interpretations.

\subsubsection{Second interpretation}

In analogy to the second interpretation proposed for the Bures metric, a
convenient coordinate system to further gain insights into the Sj\"{o}qvist
line element in Eq. (\ref{erik1}) can be achieved by performing a change
of\textbf{\ }variables defined by $p\overset{\text{def}}{=}\sin\left(
\chi\right)  $ with $\chi$ being the hyperspherical angle with $0\leq\chi
\leq\pi/2$. \ In this set of coordinates, $4ds_{\mathrm{Sj\ddot{o}qvist}}%
^{2}\left(  \vec{p}\text{, }\vec{p}+d\vec{p}\right)  $ reduces to
\begin{equation}
4ds_{\mathrm{Sj\ddot{o}qvist}}^{2}\left(  \vec{p}\text{, }\vec{p}+d\vec
{p}\right)  =d\chi^{2}+d\hat{p}\cdot d\hat{p}\text{,} \label{corre1}%
\end{equation}
with $d\hat{p}\cdot d\hat{p}=d\Omega^{2}\overset{\text{def}}{=}$ $d\theta
^{2}+\sin^{2}\left(  \theta\right)  d\varphi^{2}$. Observe that Eq.
(\ref{corre1}) for the Sj\"{o}qvist metric exhibits a structure that is
similar to that of the metric on $S^{1}\times S^{2}$, the Cartesian product of
the unit $1$-sphere $S^{1}$ with the unit $2$-sphere $S^{2}$. This Cartesian
product is responsible for the uncorrelated structure between the
hyperspherical angle coordinate and the pair of angular coordinates $\left(
\theta\text{, }\varphi\right)  $ (i.e., the polar and azimuthal angles,
respectively). This uncorrelated structure, in turn, manifests itself with an
expression of the metric on $S^{1}\times S^{2}$ which is simply the sum of the
metrics on $S^{1}$ and $S^{2}$. More specifically, comparing Eqs.
(\ref{corre0}) and (\ref{corre1}), we note that in the Sj\"{o}qvist case,
unlike the Bures case, the \textquotedblleft temporal\textquotedblright\ and
\textquotedblleft spatial\textquotedblright\ spatial components of the metric
are no longer correlated. In particular, the analogue of $\vec{x}\left(
s\right)  $ in Eq. (\ref{corre}) reduces to%
\begin{equation}
\vec{x}\left(  s\right)  =\int\frac{d\hat{p}\left(  s\right)  }{ds}ds\text{.}
\label{corre2}%
\end{equation}
From Eq. (\ref{corre2}) we emphasize the absence of a correlational structure
between the motion along $\hat{e}_{0}(s)$ and the \textquotedblleft
spatial\textquotedblright\ directions $\hat{e}_{i}(s)$ with $1\leq i\leq3$.
Interestingly, the lack of this correlational structure manifests itself when
using spherical coordinates to describe the Sj\"{o}qvist geometry.
Specifically, it emerges from the fact that the radial and angular motions
inside the Bloch sphere are not correlated since the dynamical geodesic
equations are specified by a set of second order uncoupled nonlinear
differential equations \cite{cafaroPRD22}. Lastly, recalling that line element
in the usual cylindrical coordinates $\left(  \rho\text{, }\varphi\text{,
}z\right)  $ is $ds_{\text{cylinder}}^{2}=dz^{2}+d\Omega_{\text{cylinder}}%
^{2}$ where $d\Omega_{\text{cylinder}}^{2}\overset{\text{def}}{=}d\rho
^{2}+\rho^{2}d\varphi^{2}$, we note that the structure of the Sj\"{o}qvist
line element recast as in Eq. (\ref{corre1}) is reminiscent of the structure
of a line element in the traditional cylindrical coordinates once we connect
the pair $\left(  \chi\text{, }d\Omega\right)  $ with the pair $\left(
\rho\text{, }d\Omega_{\text{cylinder}}\right)  $. Then, unlike what happens in
the Bures case, one can associate a cylinder with a constant value of its
radius in the case of the Sj\"{o}qvist geometry. In particular, the constant
value of the radius upper bounds any value that the varying radius can assume
in the Bures case. Again, as previously mentioned, these geometric insights
that arise from this simple change of coordinates would lead one to expect
different lengths of geodesic paths in the two geometries studied here.
However, this will be studied in more detail in the next section. In what
follows, instead, we present our third and last interpretation.

\subsubsection{Third interpretation}

Following the third interpretation presented for the Bures case, we adapt the
four coordinates $x^{\mu}$ with $0\leq\mu\leq3$ given by $x^{\mu}=\left(
x^{0}\text{, }\vec{x}\right)  \overset{\text{def}}{=}\left(  \sqrt{1-p^{2}%
}\text{, }\vec{p}\right)  $ with $\vec{p}=p\hat{p}$ to the Sj\"{o}qvist line
element $ds_{\mathrm{Sj\ddot{o}qvist}}^{2}\left(  \vec{p}\text{, }\vec
{p}+d\vec{p}\right)  $ in Eq. (\ref{erik1}). After some algebra, we get%
\begin{equation}
4ds_{\mathrm{Sj\ddot{o}qvist}}^{2}=\omega_{x^{0}}\left(  p\right)  \left(
dx^{0}\right)  ^{2}+\omega_{\vec{x}}\left(  p\right)  d\vec{x}\cdot d\vec
{x}\text{,} \label{buri}%
\end{equation}
with $\omega_{x^{0}}\left(  p\right)  \overset{\text{def}}{=}\left(
2p^{2}-1\right)  /p^{4}$ and $\omega_{\vec{x}}\left(  p\right)  \overset
{\text{def}}{=}1/p^{2}$. Note that Eq. (\ref{buri}) for the Sj\"{o}qvist
metric is the (extrinsic) metric for $S^{1}\times S^{2}$ embedded in $%
\mathbb{R}
^{4}$. The embedding of $S^{1}\times S^{2}$ in $%
\mathbb{R}
^{4}$ appears to be more complicated than that of $S^{3}$ in $%
\mathbb{R}
^{4}$. This complication, in turn, leads to a behavior of the Sj\"{o}qvist
metric which is more irregular than that observed in the Bures case. More
specifically, comparing Eqs. (\ref{hubi}) and (\ref{buri}), we note that
unlike what happens in the Bures case, the inside of the Bloch sphere is no
longer a unit $3$-sphere embedded in a four-dimensional flat Euclidean space
with geodesics given by great circles on it when using the four coordinates
$x^{\mu}$. In particular, the metric $4ds_{\mathrm{Sj\ddot{o}qvist}}^{2}$ as
expressed in\ Eq. (\ref{buri}) is not regular since its signature is not
constant. Indeed, $\omega_{x^{0}}\left(  p\right)  \geq0$ for $p\geq1/\sqrt
{2}$ and $\omega_{x^{0}}\left(  p\right)  \leq0$ for $0\leq p\leq1/\sqrt{2}$.
An essential singularity appears at $p=0$ (i.e., for maximally mixed states).
This observation, although obtained from a different perspective, is in
agreement with what was originally noticed in Ref. \cite{erik20}. Finally, the
geodesic paths change as well. Indeed, the geodesics $\left[  x^{\mu}\left(
s\right)  \right]  _{\mathrm{Bures}}\overset{\text{def}}{=}\left[
x^{0}\left(  s\right)  \text{, }\vec{x}\left(  s\right)  \text{ }\right]  $ in
Eq. (\ref{hubi}) are formally replaced by $\left[  x^{\mu}\left(  s\right)
\right]  _{\mathrm{Sj\ddot{o}qvist}}$ \ expressed in terms of $\left[
x^{0}\left(  s\right)  \right]  _{\mathrm{Sj\ddot{o}qvist}}$ and $\left[
\vec{x}\left(  s\right)  \right]  _{\mathrm{Sj\ddot{o}qvist}}$ as%
\begin{equation}
\left[  x^{0}\left(  s\right)  \right]  _{\mathrm{Bures}}\rightarrow\left[
x^{0}\left(  s\right)  \right]  _{\mathrm{Sj\ddot{o}qvist}}\overset
{\text{def}}{=}\int\sqrt{\left\vert \omega_{x^{0}}\left(  p\right)
\right\vert }\frac{dx^{0}\left(  s\right)  }{ds}ds\text{, } \label{me1}%
\end{equation}
and%
\begin{equation}
\left[  \vec{x}\left(  s\right)  \right]  _{\mathrm{Bures}}\rightarrow\left[
\vec{x}\left(  s\right)  \right]  _{\mathrm{Sj\ddot{o}qvist}}\overset
{\text{def}}{=}\int\frac{1}{\sqrt{\omega_{\vec{x}}\left(  p\right)  }}%
\frac{d\vec{x}\left(  s\right)  }{ds}ds\text{,} \label{me2}%
\end{equation}
respectively. For completeness and following the terminology of the previous
subsection, we point out that $p\left(  s\right)  $ in Eqs. (\ref{me1}) and
(\ref{me2}) equals $p\left(  s\right)  \overset{\text{def}}{=}\left\{
1-\left[  \cos\left(  \chi\right)  \cos\left(  s\right)  +\sin\left(
\xi\right)  \sin(s)\right]  ^{2}\right\}  ^{1/2}$ and is such that $0\leq
p(s)\leq1$.

In this section, we focused our attention on grasping physical insights from
the infinitesimal line elements for the Bures and the Sj\"{o}qvist metrics
inside a Bloch sphere in Eqs. (\ref{bures1}) and (\ref{erik1}), respectively.
In the next section, we shall further explore some of our insights by
extending our focus to the difference between the finite distances of geodesic
paths connecting mixed quantum states on these two metric manifolds.

\section{Finite distances}

In this section, we turn our attention to the study of the behaviors of the
geodesic paths on each one of the two metric manifolds, i.e. Bures and
Sj\"{o}qvist manifolds. Furthermore, we also offer a comparison between the
finite distances between arbitrary initial and final mixed states when
calculated by means of the above mentioned metrics. For clarity, we remark
that to compare finite distances between mixed quantum states in the Bloch
ball calculated with the Bures and Sj\"{o}qvist metrics, it is sufficient to
focus on points in the $xz$-plane. This is a consequence of two facts. First,
distances are preserved under rotations.\ Second, it is possible to construct
a suitable composition of two $\mathrm{SO}(3$; $%
\mathbb{R}
)$ rotations acting on arbitrary Bloch vectors for mixed states, say $\vec
{p}_{1}$ and $\vec{p}_{2}$, such that the distances $\mathcal{L}\left(
\vec{p}_{1}\text{, }\vec{p}_{2}\right)  =\mathcal{L}\left(  \vec{p}%
_{1,new}\text{, }\vec{p}_{2,new}\right)  $ with $\vec{p}_{1,new}$ and $\vec
{p}_{2,new}$ belonging to the $xz$-plane. For further details, we refer to
Appendix A.

\subsection{The Bures distance}

We begin by using spherical coordinates $\left(  r\text{, }\theta\text{,
}\varphi\right)  $ and keep $\varphi=$\textrm{const}. Then, geodesics are
obtained by minimizing $\int$ $ds_{\mathrm{Bures}}$ over all curves connecting
points $\left(  r_{a}\text{, }\theta_{a}\right)  $ and $\left(  r_{b}\text{,
}\theta_{b}\right)  $. More specifically, one arrives at the curve $\left[
\theta_{a}\text{, }\theta_{b}\right]  \ni\theta\mapsto r_{\mathrm{Bures}%
}\left(  \theta\right)  \in\left[  0\text{, }1\right]  $ that minimizes the
length $\mathcal{L}_{\mathrm{Bures}}(\vec{a}$, $\vec{b})$ defined as%
\begin{equation}
\mathcal{L}_{\mathrm{Bures}}(\vec{a}\text{, }\vec{b})\overset{\text{def}}%
{=}\int_{s_{a}}^{s_{b}}\sqrt{ds_{\mathrm{Bures}}^{2}}=\frac{1}{2}\int
_{\theta_{a}}^{\theta_{b}}\mathrm{L}\left(  r^{\prime}\text{, }r\text{,
}\theta\right)  d\theta\text{,} \label{s1a}%
\end{equation}
with $ds_{\mathrm{Bures}}^{2}$ in Eq. (\ref{burespherical}). In Eq.
(\ref{s1a}), $r^{\prime}\overset{\text{def}}{=}dr/d\theta$ and $\mathrm{L}%
\left(  r^{\prime}\text{, }r\text{, }\theta\right)  $ is the Lagrangian-like
function defined as%
\begin{equation}
\mathrm{L}\left(  r^{\prime}\text{, }r\text{, }\theta\right)  \overset
{\text{def}}{=}\sqrt{r^{2}+\frac{r^{\prime2}}{1-r^{2}}}\text{.} \label{s2a}%
\end{equation}
From Eq. (\ref{s2}), note that $\mathrm{L}=\mathrm{L}\left(  r^{\prime}\text{,
}r\right)  $ in Eq.\ (\ref{s2a}) does not explicitly depend $\theta$.
Therefore, $\partial\mathrm{L}/\partial\theta=0$. In this case, it happens
that the Euler-Lagrange equation%
\begin{equation}
\frac{d}{d\theta}\frac{\partial\mathrm{L}\left(  r^{\prime}\text{, }r\right)
}{\partial r^{\prime}}-\frac{\partial\mathrm{L}\left(  r^{\prime}\text{,
}r\right)  }{\partial r}=0\text{,} \label{EL1}%
\end{equation}
reduces to the well-known Beltrami identity in Lagrangian mechanics,%
\begin{equation}
\mathrm{L}\left(  r^{\prime}\text{, }r\right)  -r^{\prime}\frac{\partial
\mathrm{L}\left(  r^{\prime}\text{, }r\right)  }{\partial r^{\prime}%
}=\text{\textrm{const}.} \label{Beltrami1}%
\end{equation}
Indeed, making use of Eq. (\ref{EL1}) together with the identity%
\begin{equation}
\frac{d\mathrm{L}\left(  r^{\prime}\text{, }r\right)  }{d\theta}%
=\frac{\partial\mathrm{L}\left(  r^{\prime}\text{, }r\right)  }{\partial
r^{\prime}}r^{\prime\prime}+\frac{\partial\mathrm{L}\left(  r^{\prime}\text{,
}r\right)  }{\partial r}r^{\prime}\text{,}%
\end{equation}
we get%
\begin{equation}
\frac{d\mathrm{L}\left(  r^{\prime}\text{, }r\right)  }{d\theta}=\frac
{d}{d\theta}\left(  r^{\prime}\frac{\partial\mathrm{L}\left(  r^{\prime
}\text{, }r\right)  }{\partial r^{\prime}}\right)  \text{.} \label{bebe}%
\end{equation}
Finally, Eq. (\ref{bebe}) leads to the so-called Beltrami identity in Eq.
(\ref{Beltrami1}). Using Eqs. (\ref{s2a}) and (\ref{Beltrami1}), we
obtain\ \ \ \ \
\begin{equation}
\frac{r^{2}}{\sqrt{r^{2}+\frac{r^{\prime2}}{1-r^{2}}}}=\text{\textrm{const}%
.}\equiv\mathrm{c}_{\mathrm{B}}\text{.} \label{ca}%
\end{equation}
Manipulating Eq. (\ref{ca}) and imposing the boundary conditions $r\left(
\theta_{a}\right)  =r_{a}$ and $r\left(  \theta_{b}\right)  =r_{b}$, we obtain%
\begin{equation}
\int_{r_{a}}^{r_{b}}\frac{dr}{\sqrt{r^{2}\left(  \frac{r^{2}}{\mathrm{c}%
_{\mathrm{B}}^{2}}-1\right)  \left(  1-r^{2}\right)  }}=\int_{\theta_{a}%
}^{\theta_{b}}d\theta\text{,} \label{ca1}%
\end{equation}
with $0<\mathrm{c}_{\mathrm{B}}\leq r\leq1$. For notational simplicity, let us
set $\mathrm{a}_{\mathrm{B}}^{2}\overset{\text{def}}{=}1/\mathrm{c}%
_{\mathrm{B}}^{2}>1$. Then, integration of Eq. (\ref{ca1}) by use of
Mathematica yields%
\begin{equation}
\mathrm{I}\left(  r\right)  \overset{\text{def}}{=}\int\frac{dr}{\sqrt
{r^{2}\left(  \frac{r^{2}}{\mathrm{c}_{\mathrm{B}}^{2}}-1\right)  \left(
1-r^{2}\right)  }}=\frac{r\sqrt{r^{2}-1}\sqrt{\mathrm{a}_{\mathrm{B}}^{2}%
r^{2}-1}}{\sqrt{-r^{2}\left(  r^{2}-1\right)  \left(  \mathrm{a}_{\mathrm{B}%
}^{2}r^{2}-1\right)  }}\tanh^{-1}\left(  \frac{\sqrt{r^{2}-1}}{\sqrt
{\mathrm{a}_{\mathrm{B}}^{2}r^{2}-1}}\right)  +\text{\textrm{const}.}
\label{ca2}%
\end{equation}
Manipulating Eq. (\ref{ca2}), $\mathrm{I}\left(  r\right)  $ in Eq.
(\ref{ca2}) becomes%
\begin{equation}
\mathrm{I}\left(  r\right)  =-\arctan\left(  \frac{\sqrt{1-r^{2}}}%
{\sqrt{\mathrm{a}_{\mathrm{B}}^{2}r^{2}-1}}\right)  +\text{\textrm{const}.}
\label{ca3}%
\end{equation}
Finally, substituting Eq. (\ref{ca3}) into Eq. (\ref{ca1}), the radial
geodesic path in the Bures case can be recast as \cite{cafaroPRD22}%
\begin{equation}
r_{\mathrm{Bures}}\left(  \theta\right)  =\sqrt{\frac{1+\tan^{2}\left[
\mathcal{A}_{\mathrm{B}}\left(  r_{a}\text{, }r_{a}^{\prime}\right)  -\left(
\theta-\theta_{a}\right)  \right]  }{1+\mathrm{a}_{\mathrm{B}}^{2}\left(
r_{a}\text{, }r_{a}^{\prime}\right)  \tan^{2}\left[  \mathcal{A}_{\mathrm{B}%
}\left(  r_{a}\text{, }r_{a}^{\prime}\right)  -\left(  \theta-\theta
_{a}\right)  \right]  }}\text{,} \label{ca4}%
\end{equation}
where the constants $\mathcal{A}_{\mathrm{B}}\left(  r_{a}\text{, }%
r_{a}^{\prime}\right)  $ and $\mathrm{a}_{\mathrm{B}}^{2}\left(  r_{a}\text{,
}r_{a}^{\prime}\right)  $ in Eq. (\ref{ca4}) are given by%
\begin{equation}
\mathcal{A}_{\mathrm{B}}\left(  r_{a}\text{, }r_{a}^{\prime}\right)
\overset{\text{def}}{=}\arctan\left(  \sqrt{\left(  1-r_{a}^{2}\right)
^{2}(\frac{r_{a}}{r_{a}^{\prime}})^{2}}\right)  \text{, and }\mathrm{a}%
_{\mathrm{B}}^{2}\left(  r_{a}\text{, }r_{a}^{\prime}\right)  \overset
{\text{def}}{=}\frac{1}{r_{a}^{2}}+(\frac{r_{a}^{\prime}}{r_{a}})^{2}\frac
{1}{r_{a}^{2}\left(  1-r_{a}^{2}\right)  }\text{,}%
\end{equation}
respectively. At this point, we recall that the Bures distance between two
density operators $\rho_{1}$ and $\rho_{2}$ is given by \cite{hubner92},%
\begin{equation}
\mathcal{L}_{\mathrm{Bures}}\left(  \rho_{1},\rho_{2}\right)  \overset
{\text{def}}{=}\sqrt{2}\left[  1-\mathrm{tr}\left(  \sqrt{\sqrt{\rho_{1}}%
\rho_{2}\sqrt{\rho_{1}}}\right)  \right]  ^{1/2}\text{.} \label{BD1}%
\end{equation}
Furthermore, $\mathcal{L}_{\mathrm{Bures}}\left(  \rho_{1},\rho_{2}\right)  $
in\ Eq. (\ref{BD1}) can be expressed in terms of the fidelity between two
density operators $\rho_{1}$ and $\rho_{2}$ defined as \cite{jozsa94,wilde17}%
\begin{equation}
\mathrm{F}\left(  \rho_{1},\rho_{2}\right)  \overset{\text{def}}{=}\left(
\mathrm{tr}\left(  \sqrt{\sqrt{\rho_{1}}\rho_{2}\sqrt{\rho_{1}}}\right)
\right)  ^{2}\text{.} \label{BD2}%
\end{equation}
Therefore, combining Eqs. (\ref{BD1}) and (\ref{BD2}), the Bures distance
$\mathcal{L}_{\mathrm{Bures}}\left(  \rho_{1},\rho_{2}\right)  $ becomes%
\begin{equation}
\mathcal{L}_{\mathrm{Bures}}\left(  \rho_{1},\rho_{2}\right)  =\sqrt{2}\left[
1-\sqrt{\mathrm{F}\left(  \rho_{1},\rho_{2}\right)  }\right]  ^{1/2}\text{.}
\label{BD3}%
\end{equation}
Then, focusing on qubit states, the fidelity $\mathrm{F}\left(  \rho_{1}%
,\rho_{2}\right)  $ in Eq. (\ref{BD2}) reduces to \cite{jozsa94}%
\begin{equation}
\mathrm{F}\left(  \rho_{1},\rho_{2}\right)  =\mathrm{tr}\left(  \rho_{1}%
\rho_{2}\right)  +2\sqrt{\det\left(  \rho_{1}\right)  \det\left(  \rho
_{2}\right)  }\text{,} \label{BD4}%
\end{equation}
where $\rho_{1}$ and $\rho_{2}$ are given by%
\begin{equation}
\rho_{1}=\rho\left(  \vec{a}\right)  \overset{\text{def}}{=}\frac
{\mathrm{I}+\vec{a}\cdot\vec{\sigma}}{2}\text{, and }\rho_{2}=\rho(\vec
{b})\overset{\text{def}}{=}\frac{\mathrm{I}+\vec{b}\cdot\vec{\sigma}}%
{2}\text{,} \label{BD5}%
\end{equation}
respectively. Using Eqs. (\ref{BD5}) and (\ref{BD4}), $\mathcal{L}%
_{\mathrm{Bures}}\left(  \rho_{1},\rho_{2}\right)  $ in Eq. (\ref{BD3})
reduces to%
\begin{equation}
\mathcal{L}_{\mathrm{Bures}}\left(  \rho_{1},\rho_{2}\right)  =\sqrt{2}\left[
1-\sqrt{2\left(  \frac{1+\vec{a}\cdot\vec{b}}{4}+\sqrt{\frac{1-\vec{a}^{2}}%
{4}\frac{1-\vec{b}^{2}}{4}}\right)  }\right]  ^{1/2}\text{.} \label{BD6}%
\end{equation}
For Bloch vectors $\vec{a}\overset{\text{def}}{=}r_{a}\hat{n}_{a}$ and
$\vec{b}\overset{\text{def}}{=}r_{b}\hat{n}_{b}$ in the $xz$-plane, we have
$\hat{n}_{a}\overset{\text{def}}{=}\left(  \sin\left(  \theta_{a}\right)
\text{, }0\text{, }\cos\left(  \theta_{a}\right)  \right)  $ and $\hat{n}%
_{b}\overset{\text{def}}{=}\left(  \sin\left(  \theta_{b}\right)  \text{,
}0\text{, }\cos\left(  \theta_{b}\right)  \right)  $ with $\theta_{a}$ and
$\theta_{b}$ in $\left[  0,\pi\right]  $. Therefore, in this scenario,
$\mathcal{L}_{\mathrm{Bures}}\left(  \rho_{1},\rho_{2}\right)  =\mathcal{L}%
_{\mathrm{Bures}}(\rho_{1}\left(  \vec{a}\right)  ,\rho_{2}(\vec
{b}))=\mathcal{L}_{\mathrm{Bures}}(\vec{a}$, $\vec{b})$ in Eq. (\ref{BD6}) is
equal to%
\begin{equation}
\mathcal{L}_{\mathrm{Bures}}(\vec{a}\text{, }\vec{b})=\sqrt{2}\left[
1-\sqrt{2\left(  \frac{1+r_{a}r_{b}\cos\left(  \theta_{a}-\theta_{b}\right)
}{4}+\sqrt{\frac{1-r_{a}^{2}}{4}\frac{1-r_{b}^{2}}{4}}\right)  }\right]
^{1/2}\text{.} \label{BD7}%
\end{equation}
The expression of $\mathcal{L}_{\mathrm{Bures}}(\vec{a}$, $\vec{b})$ in Eq.
(\ref{BD7}) helps us evaluating the finite distance between arbitrary mixed
states $\rho\left(  \vec{a}\right)  \overset{\text{def}}{=}(\mathrm{I}+\vec
{a}\cdot\vec{\sigma})/2$ and $\rho(\vec{b})\overset{\text{def}}{=}%
(\mathrm{I}+\vec{b}\cdot\vec{\sigma})/2$ belonging to the $xz$-plane and,
thus, for arbitrary states in the Bloch ball (for details, see Appendix
A).\begin{figure}[t]
\centering
\includegraphics[width=0.4\textwidth] {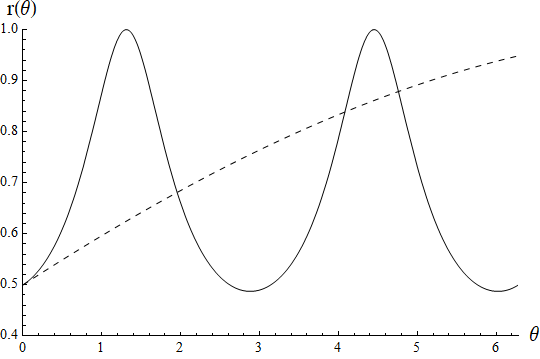}\caption{Illustrative depiction of
Bures and Sj\"{o}qvist curves $r_{\mathrm{Bures}}\left(  \theta\right)  $
(solid line) and $r_{\mathrm{Sjoqvist}}\left(  \theta\right)  $ (dashed line),
respectively, for identical boundary conditions specified by the initial
position ($r\left(  \theta_{a}\right)  =r_{a}\overset{\text{def}}{=}1/2$) and
the initial speed ($r^{\prime}\left(  \theta_{a}\right)  =r_{a}^{\prime
}\overset{\text{def}}{=}0.1$). Note that $\theta_{a}\leq\theta\leq\theta_{b}$,
with $\theta_{a}\overset{\text{def}}{=}0$ and $\theta_{b}\overset{\text{def}%
}{=}2\pi$.}%
\end{figure}

\subsection{The Sj\"{o}qvist distance}

Following Sj\"{o}qvist's work in Ref. \cite{erik20}, we focus on finding
geodesic paths that connect points (i.e., mixed quantum states) in the Bloch
ball that lay in a plane that contains the origin. Employing spherical
coordinates $\left(  r\text{, }\theta\text{, }\varphi\right)  $ and
maintaining $\varphi=$\textrm{const.}, geodesics can be obtained by minimizing
$\int$ $ds_{\mathrm{Sj\ddot{o}qvist}}$ over all curves connecting points
$\vec{a}\leftrightarrow\left(  r_{a}\text{, }\theta_{a}\right)  $ and $\vec
{b}\leftrightarrow\left(  r_{b}\text{, }\theta_{b}\right)  $. More
specifically, we aim to get the curve $\left[  \theta_{a}\text{, }\theta
_{b}\right]  \ni\theta\mapsto r_{\mathrm{Sj\ddot{o}qvist}}\left(
\theta\right)  \in\left(  0\text{, }1\right]  $ that minimizes the length
$\mathcal{L}_{\mathrm{Sj\ddot{o}qvist}}(\vec{a}$, $\vec{b})$ given by%
\begin{equation}
\mathcal{L}_{\mathrm{Sj\ddot{o}qvist}}(\vec{a}\text{, }\vec{b})\overset
{\text{def}}{=}\int_{s_{a}}^{s_{b}}\sqrt{ds_{\mathrm{Sj\ddot{o}qvist}}^{2}%
}=\frac{1}{2}\int_{\theta_{a}}^{\theta_{b}}\mathrm{L}\left(  r^{\prime}\text{,
}r\text{, }\theta\right)  d\theta\text{,} \label{s1}%
\end{equation}
with $ds_{\mathrm{Sj\ddot{o}qvist}}^{2}$ in Eq. (\ref{sspherical}). In Eq.
(\ref{s1}), $r^{\prime}\overset{\text{def}}{=}dr/d\theta$ and $\mathrm{L}%
\left(  r^{\prime}\text{, }r\text{, }\theta\right)  $ is the Lagrangian-like
function defined as%
\begin{equation}
\mathrm{L}\left(  r^{\prime}\text{, }r\text{, }\theta\right)  \overset
{\text{def}}{=}\sqrt{1+\frac{r^{\prime2}}{1-r^{2}}}\text{.} \label{s2}%
\end{equation}
Following the similar derivation of the Euler-Lagrange equations leading
to\ Eq. (\ref{ca}) in the previous subsection for the Bures metric, we
obtain\textbf{ } \ \ \
\begin{equation}
\frac{1}{\sqrt{1+\frac{r^{\prime2}}{1-r^{2}}}}=\text{\textrm{const}.}%
\equiv\mathrm{c}_{\mathrm{S}}\text{,} \label{c}%
\end{equation}
that is,%
\begin{equation}
\frac{r^{\prime2}}{1-r^{2}}=\text{\textrm{const}.}\equiv\mathrm{k}%
\overset{\text{def}}{=}\frac{1-\mathrm{c}_{\mathrm{S}}^{2}}{\mathrm{c}%
_{\mathrm{S}}^{2}}\text{,} \label{be2}%
\end{equation}
Integrating Eq. (\ref{be2}) and imposing the boundary conditions $r\left(
\theta_{a}\right)  =r_{a}$, $r\left(  \theta_{b}\right)  =r_{b}$, with
$\theta_{\mathrm{initial}}=\theta_{a}$ and $\theta_{\mathrm{final}}=\theta
_{b}$, we obtain the geodesic path%
\begin{equation}
r_{\mathrm{Sj\ddot{o}qvist}}\left(  \theta\right)  =\sin\left\{  \frac
{\sin^{-1}\left(  r_{b}\right)  -\sin^{-1}\left(  r_{a}\right)  }{\theta
_{b}-\theta_{a}}\theta+\frac{\sin^{-1}\left(  r_{a}\right)  +\sin^{-1}\left(
r_{b}\right)  }{2}-\frac{1}{2}\frac{\theta_{a}+\theta_{b}}{\theta_{b}%
-\theta_{a}}\left[  \sin^{-1}\left(  r_{b}\right)  -\sin^{-1}\left(
r_{a}\right)  \right]  \right\}  \text{.} \label{rerik}%
\end{equation}
We remark that the expression of $r_{\mathrm{Sj\ddot{o}qvist}}\left(
\theta\right)  $ in Eq. (\ref{rerik}) can be recast, alternatively, in terms
of boundary conditions on the initial position $r\left(  \theta_{a}\right)
=r_{a}$ and the initial speed $r^{\prime}\left(  \theta_{a}\right)
=r_{a}^{\prime}$. We get, after some algebra \cite{cafaroPRD22},%
\begin{equation}
r_{\mathrm{Sj\ddot{o}qvist}}\left(  \theta\right)  =\sin\left[  \frac
{r_{a}^{\prime}}{\sqrt{1-r_{a}^{2}}}\left(  \theta-\theta_{a}\right)
+\sin^{-1}\left(  r_{a}\right)  \right]  \text{.} \label{rerikGOOD}%
\end{equation}
As a consistency check, we observe that we correctly recover Eq. (18) in Ref.
\cite{erik20} when we set $\theta_{\mathrm{initial}}=0$ in Eq. (\ref{rerik}).
For illustrative purposes, we present in Fig. $1$ a plot of the Bures and
Sj\"{o}qvist curves $r_{\mathrm{Bures}}\left(  \theta\right)  $ in Eq.
(\ref{ca4}) and $r_{\mathrm{Sjoqvist}}\left(  \theta\right)  $ in Eq.
(\ref{rerikGOOD}), respectively, for identical boundary conditions specified
by the initial position and the initial speed. Finally, inserting
$r_{\mathrm{Sj\ddot{o}qvist}}\left(  \theta\right)  $ in Eq. (\ref{rerik})
into the expression for $\mathrm{L}\left(  r^{\prime}\text{, }r\text{, }%
\theta\right)  $ in Eq. (\ref{s2}), $\mathcal{L}_{\mathrm{Sj\ddot{o}qvist}%
}(\vec{a}$, $\vec{b})$ in Eq. (\ref{s1}) reduces to \cite{erik20}%
\begin{equation}
\mathcal{L}_{\mathrm{Sj\ddot{o}qvist}}(\vec{a}\text{, }\vec{b})=\frac{1}%
{2}\sqrt{\left(  \theta_{b}-\theta_{a}\right)  ^{2}+\left[  \sin^{-1}\left(
r_{b}\right)  -\sin^{-1}\left(  r_{a}\right)  \right]  ^{2}}\text{.}
\label{asslength}%
\end{equation}
The expression of $\mathcal{L}_{\mathrm{Sj\ddot{o}qvist}}(\vec{a}$, $\vec{b})$
in Eq. (\ref{asslength}) allows us to calculate the finite distance between
arbitrary mixed states $\rho\left(  \vec{a}\right)  \overset{\text{def}}%
{=}(\mathrm{I}+\vec{a}\cdot\vec{\sigma})/2$ and $\rho(\vec{b})\overset
{\text{def}}{=}(\mathrm{I}+\vec{b}\cdot\vec{\sigma})/2$ laying in the
$xz$-plane and, thus, for arbitrary states in the Bloch ball (for details, see
Appendix A). For illustrative purposes, we plot in part (a) of Fig. $2$ the
Sj\"{o}qvist distance $\mathcal{L}_{\mathrm{Sj\ddot{o}qvist}}\left(
\Delta\theta\right)  $ in\ Eq. (\ref{asslength}) and the Bures distance
$\mathcal{L}_{\mathrm{Bures}}\left(  \Delta\theta\right)  $ in Eq. (\ref{BD7})
versus $\Delta\theta\overset{\text{def}}{=}\theta_{b}-\theta_{a}$ with
$0\leq\Delta\theta\leq\pi$ in the assumption that $r_{a}=r_{b}=1$. In part (b)
of Fig. $2$, instead, we compare the Sj\"{o}qvist and Bures distances as in
(a) but for different values of $r_{a}=r_{b}=$\textrm{const. }with
\textrm{const.}$\in\left\{  1\text{, }0.95\text{, }0.75\text{, }0.5\right\}
$. We observe that while the Sj\"{o}qvist distance does not depend on the
particular value of the \textrm{const.}, the Bures distance depends on the
specific value of the \textrm{const.} In any case, we have $0\leq$
$\mathcal{L}_{\mathrm{Bures}}\left(  \Delta\theta\right)  \leq\mathcal{L}%
_{\mathrm{Sj\ddot{o}qvist}}\left(  \Delta\theta\right)  \leq\pi/2$. For
completeness, we recall that the geodesic distance between two orthogonal pure
states represented by antipodal points on the Bloch sphere is $\pi$, whereas
the corresponding Fubini-Study distance is $\pi/2$. In the limit of
$r_{a}=r_{b}=1$, the Bures distance in Eq. (\ref{BD7}) reduces to the
Fubini-Study distance $\left\vert \Delta\theta\right\vert /2$ with
$\Delta\theta\overset{\text{def}}{=}\theta_{b}-\theta_{a}$ only approximately
since $\mathcal{L}_{\mathrm{Bures}}\left(  \Delta\theta\right)  =\left\vert
\Delta\theta\right\vert /2+O\left(  \left\vert \Delta\theta\right\vert
^{3}\right)  $ when $\left\vert \Delta\theta\right\vert \ll1$. In the same
limiting case of $r_{a}=r_{b}=1$, instead, the Sj\"{o}qvist distance in Eq.
(\ref{asslength}) reduces to the Fubini-Study distance $\left\vert
\Delta\theta\right\vert /2$ in an exact manner.

Thanks to Eqs. (\ref{BD7}) and (\ref{asslength}), we are now ready to provide
some intriguing discussion points in the next section.\begin{figure}[t]
\centering
\includegraphics[width=1\textwidth] {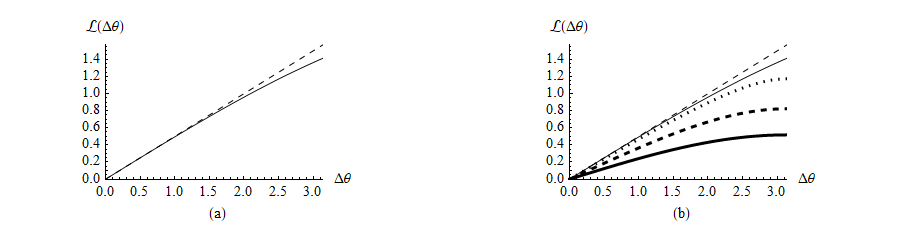}\caption{In (a), we plot the
Sj\"{o}qvist distance $\mathcal{L}_{\mathrm{Sj\ddot{o}qvist}}\left(
\Delta\theta\right)  $ (dashed line) and the Bures distance $\mathcal{L}%
_{\mathrm{Bures}}\left(  \Delta\theta\right)  $ (solid line) versus
$\Delta\theta\overset{\text{def}}{=}\theta_{b}-\theta_{a}$ with $0\leq
\Delta\theta\leq\pi$ in the assumption that $r_{a}=r_{b}=1$. In (b), we
compare the Sj\"{o}qvist and Bures distances as in (a) but for different
values of $r_{a}=r_{b}=$\textrm{const. }with \textrm{const.}$\in\left\{
1\text{, }0.95\text{, }0.75\text{, }0.5\right\}  $. While the Sj\"{o}qvist
distance (thin dashed line) does not depend on the particular value of the
\textrm{const.}, the Bures distance depends on the specific value of the
\textrm{const. }(thin solid fine for $r=1$, thick dotted line for $r=0.95$,
thick dashed line for $r=0.75$, and thick solid line for $r=0.5$). In any
case, we have $0\leq$ $\mathcal{L}_{\mathrm{Bures}}\left(  \Delta
\theta\right)  \leq\mathcal{L}_{\mathrm{Sj\ddot{o}qvist}}\left(  \Delta
\theta\right)  \leq\pi/2\approx1.57$.}%
\end{figure}\begin{figure}[t]
\centering
\includegraphics[width=0.6\textwidth] {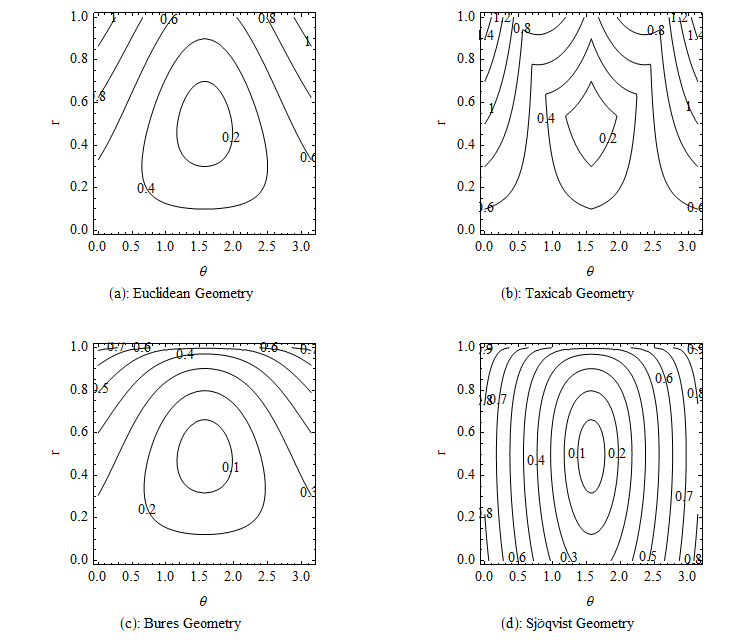}\caption{In (a), we illustrate the
Euclidean geometry in terms of a contour plot that exhibits the spherical
coordinates $r$ v.s. $\theta$ with $0\leq r\leq1$ and $0\leq\theta\leq\pi$.
The level curves are given by $\mathcal{L}_{\mathrm{Euclid}}(\vec{a}$,
$\vec{b})=c$, with $c$ being a positive constant, $\vec{a}\overset{\text{def}%
}{=}(r_{a}\cos\left(  \theta_{a}\right)  $, $r_{a}\sin\left(  \theta
_{a}\right)  )=\left(  0\text{, }1/2\right)  $ with $\left(  r_{a}\text{,
}\theta_{a}\right)  \overset{\text{def}}{=}(1/2$, $\pi/2)$ , and $\vec
{b}\overset{\text{def}}{=}\left(  r\cos(\theta)\text{, }r\sin(\theta)\right)
$. In (b), we follow (a) and depict the Taxicab geometry with level curves
specified by the relation $\mathcal{L}_{\mathrm{Taxicab}}(\vec{a}$, $\vec
{b})=c$. In (c), we illustrate the Bures geometry in terms of the contour plot
that shows the spherical coordinates $r$ v.s. $\theta$ with $0\leq r\leq1$ and
$0\leq\theta\leq\pi$. The level curves are characterized by the relation
$\mathcal{L}_{\mathrm{Bures}}(\vec{a}$, $\vec{b})=c$, with $c$ being a
positive constant. The vectors $\vec{a}$ and $\vec{b}$ are as in (a) and (b).
Finally, following (c), we depict in (d) the Sj\"{o}qvist geometry with level
curves defined by $\mathcal{L}_{\mathrm{Sj\ddot{o}qvist}}(\vec{a}$, $\vec
{b})=c$. Again, the vectors $\vec{a}$ and $\vec{b}$ are as in (a), (b), and
(c). Finally, the expressions for $\mathcal{L}_{\mathrm{Euclid}}(\vec{a}$,
$\vec{b})$, $\mathcal{L}_{\mathrm{Taxicab}}(\vec{a}$, $\vec{b})$,
$\mathcal{L}_{\mathrm{Bures}}(\vec{a}$, $\vec{b})$, and $\mathcal{L}%
_{\mathrm{Sj\ddot{o}qvist}}(\vec{a}$, $\vec{b})$ are the ones that appear in
the main text.}%
\end{figure}

\section{Discussion}

To better motivate and understand the relevance of our forthcoming discussion,
we briefly summarize some of the main results we found in past investigations
on a comparative analysis of the Bures and Sjoqvist metrics. In Ref.
\cite{cafaroPRD22}, we found that the manifold of mixed states equipped with
the Bures (Sj\"{o}qvist) metric is an isotropic (anisotropic) manifold of
constant (non-constant) sectional curvature. The isotropy of the manifold, the
inequality $\mathcal{L}_{\mathrm{Bures}}(\vec{a},\vec{b})\leq$ $\mathcal{L}%
_{\mathrm{Sj\ddot{o}qvist}}(\vec{a},\vec{b})$ between the path lengths and, in
addition, the presence of a correlational structure in the equations of
geodesic motion (which is absent in the Sj\"{o}qvist case) between radial and
angular directions, are at the root of the softening in the complexity of the
geodesic evolution on Bures manifolds. Indeed, correlational structures cause
the shrinkage of the explored volumes of regions on the manifold underlying
the geodesic evolution. This shrinkage, finally, can be detected by means of
the so-called information geometric complexity (i.e., the volume of the
parametric region explored by the system during its evolution from the initial
to the final configuration on the underlying manifold, \cite{cafarothesis}%
).\textbf{ }For a summary of the specific properties of Bures and Sj\"{o}qvist
metrics in terms of sectional curvatures, path lengths, and information
geometric complexities, we refer to Table III and Appendix E in Ref.
\cite{cafaroPRD22}. We also point out that we originally observed in Ref.
\cite{cafaro23EPJ} that the Bures and Sj\"{o}qvist metrics characterize, in
general, the departure from the classical behavior by means of the
noncommutativity of neighboring mixed states in dissimilar manners. This
discrepancy was first tested by studying geometric aspects of the Bures and
Sj\"{o}qvist manifolds emerging from a superconducting flux Hamiltonian model
in Ref. \cite{cafaro23EPJ}. Later, this discrepancy was elegantly
conceptualized (see Eqs. (36) and (38) in Ref. \cite{alsing23}) and, in
addition, explicitly discussed for a spin-qubit in an arbitrarily oriented
uniform and stationary magnetic field in thermal equilibrium with a
finite-temperature reservoir in Ref. \cite{alsing23}.

In this section, we briefly comment on some previously unnoticed geometric
features that emerge from the Bures and Sj\"{o}qvist finite distances in Eqs.
(\ref{BD7}) and (\ref{asslength}) obtained in Section III. To make our
discussion closer to classical geometric and topological arguments, we carry
out a comparative discussion highlighting formal similarities between the
classical (Euclidean, Taxicab) metrics in the\textbf{ }$xz$\textbf{-}plane
of\textbf{ }$%
\mathbb{R}
^{2}$ and the quantum (Bures, Sj\"{o}qvist) metrics inside the Bloch sphere.
Let us denote with $d_{\mathrm{Euclid}}$ and $d_{\mathrm{Taxicab}}$ the usual
Euclidean and Taxicab metric functions, respectively.\textbf{ }For
completeness, we recall that $d_{\mathrm{Euclid}}(\vec{a},\vec{b}%
)\overset{\text{def}}{=}\sqrt{\left(  a_{x}-b_{x}\right)  ^{2}+\left(
a_{z}-b_{z}\right)  ^{2}}$ and $d_{\mathrm{Taxicab}}(\vec{a},\vec{b}%
)\overset{\text{def}}{=}\left\vert a_{x}-b_{x}\right\vert +\left\vert
a_{z}-b_{z}\right\vert $ with $\vec{a}\overset{\text{def}}{=}\left(
a_{x},a_{z}\right)  ,\vec{b}\overset{\text{def}}{=}\left(  b_{x},b_{z}\right)
$ in $%
\mathbb{R}
^{2}$. First, we observe that although $\left(
\mathbb{R}
^{2}\text{, }d_{\mathrm{Euclid}}\right)  $ and $\left(
\mathbb{R}
^{2}\text{, }d_{\mathrm{Taxicab}}\right)  $ are topologically equivalent
metric spaces \cite{croom23}, we have that $d_{\mathrm{Euclid}}\left(
P_{i}\text{, }P_{i^{\prime}}\right)  \leq d_{\mathrm{Euclid}}\left(
P_{k}\text{, }P_{k^{\prime}}\right)  $ does not imply that
$d_{\mathrm{Taxicab}}\left(  P_{i}\text{, }P_{i^{\prime}}\right)  \leq
d_{\mathrm{Taxicab}}\left(  P_{k}\text{, }P_{k^{\prime}}\right)  $ with $i\neq
i^{\prime}$ and $k\neq k^{\prime}$. Therefore, a relative ranking of pairs of
points specified in terms of distances between the pairs themselves, with
closer pairs of points ranking higher than those further away, is not
preserved when using Euclidean and Taxicab metrics. For instance, consider a
set $\mathcal{S}_{1}$ of three points in $%
\mathbb{R}
^{2}$ given in Cartesian coordinates by $\mathcal{S}_{1}\overset{\text{def}%
}{=}\left\{  P_{1}\overset{\text{def}}{=}\left(  0,0\right)  \text{, }%
P_{2}\overset{\text{def}}{=}\left(  0,1\right)  \text{, }P_{3}\overset
{\text{def}}{=}((1+\sqrt{2})/4,(1+\sqrt{2})/4)\right\}  $. One notices that
$d_{\mathrm{Euclid}}\left(  P_{1}\text{, }P_{2}\right)  =1\geq0.85\simeq
d_{\mathrm{Euclid}}\left(  P_{1}\text{, }P_{3}\right)  $. However, when using
the Taxicab metric, we have $d_{\mathrm{Taxicab}}\left(  P_{1}\text{, }%
P_{2}\right)  =1\leq1.21\simeq d_{\mathrm{Taxicab}}\left(  P_{1}\text{, }%
P_{3}\right)  $. Interestingly, the conservation of this type of ranking of
pairs of points is violated also when comparing the Bures and Sj\"{o}qvist
metrics. For instance, consider a set $\mathcal{S}_{2}$ of four points (i.e.,
mixed quantum states) $P_{i}\leftrightarrow\rho\left(  P_{i}\right)
=(\mathrm{I}+\vec{a}_{i}\cdot\vec{\sigma})/2$ with $\vec{a}_{i}$ assumed to
belong to the $xz$-plane and specified by the pair of spherical coordinates
$\left(  r_{a_{i}}\text{, }\theta_{a_{i}}\right)  $ given by $\mathcal{S}%
_{2}\overset{\text{def}}{=}\left\{  P_{1}\overset{\text{def}}{=}\left(
1/2,0\right)  \text{, }P_{2}\overset{\text{def}}{=}\left(  1/2,\pi\right)
\text{, }P_{3}\overset{\text{def}}{=}(1/8,0)\text{, }P_{4}\overset{\text{def}%
}{=}(1/4,\pi)\right\}  $. Then, in terms of the Bures metric, we find
$d_{\mathrm{Bures}}\left(  P_{1}\text{, }P_{2}\right)  \simeq0.52\geq
0.19\simeq d_{\mathrm{Bures}}\left(  P_{3}\text{, }P_{4}\right)  $. However,
when using the Sj\"{o}qvist metric, we get $d_{\mathrm{Sj\ddot{o}qvist}%
}\left(  P_{1}\text{, }P_{2}\right)  =\pi/2\simeq1.570\leq1.572\simeq
d_{\mathrm{Sj\ddot{o}qvist}}\left(  P_{3}\text{, }P_{4}\right)  $. Clearly,
$d_{\mathrm{Bures}}\left(  P_{i}\text{, }P_{j}\right)  =d_{\mathrm{Bures}%
}\left(  \vec{a}_{i}\text{, }\vec{a}_{j}\right)  =\mathcal{L}_{\mathrm{Bures}%
}\left(  \vec{a}_{i}\text{, }\vec{a}_{j}\right)  $ as defined in Eq.
(\ref{BD7}). Similarly, $d_{\mathrm{Sj\ddot{o}qvist}}\left(  P_{i}\text{,
}P_{j}\right)  =d_{\mathrm{Sj\ddot{o}qvist}}\left(  \vec{a}_{i}\text{, }%
\vec{a}_{j}\right)  =\mathcal{L}_{\mathrm{Sj\ddot{o}qvist}}\left(  \vec{a}%
_{i}\text{, }\vec{a}_{j}\right)  $ as defined in Eq. (\ref{asslength}). We
also emphasize here that unlike what happens in the Bures geometry, in the
Sj\"{o}qvist geometry it is possible to identify pairs of two points, say
$\left(  P_{i}\text{, }P_{i^{\prime}}\right)  $ and $\left(  P_{k}\text{,
}P_{k^{\prime}}\right)  $, that seem to be visually rankable which, in
actuality, are at the same distance from each other (and, thus, non-rankable
according to our previously mentioned notion of relative ranking). For
example, following the terminology introduced for the set $\mathcal{S}_{2}$,
consider the new set of points $\mathcal{S}_{3}$ defined as $\mathcal{S}%
_{3}\overset{\text{def}}{=}\left\{  P_{1}\overset{\text{def}}{=}\left(
1/4,0\right)  \text{, }P_{2}\overset{\text{def}}{=}\left(  1/4,\pi\right)
\text{, }P_{3}\overset{\text{def}}{=}(1/2,0)\text{, }P_{4}\overset{\text{def}%
}{=}(1/2,\pi)\right\}  $. Then, when employing the Sj\"{o}qvist metric, we
find $d_{\mathrm{Sj\ddot{o}qvist}}\left(  P_{1}\text{, }P_{2}\right)
=d_{\mathrm{Sj\ddot{o}qvist}}\left(  P_{3}\text{, }P_{4}\right)  =\pi/2$, even
though the pair of points $\left(  P_{3}\text{, }P_{4}\right)  $ seem to be
visually more distant than the pair of points $\left(  P_{1}\text{, }%
P_{2}\right)  $. However, when employing the Bures metric, we get
$d_{\mathrm{Bures}}\left(  P_{1}\text{, }P_{2}\right)  \simeq0.25\leq
0.52\simeq d_{\mathrm{Bures}}\left(  P_{3}\text{, }P_{4}\right)  $. This
latter inequality is consistent with our visual intuition associated with
seeing these points as mixed states inside the Bloch sphere. Clearly, these
different geometric features between Bures and Sj\"{o}qvist geometries can be
ascribed to the formal structure of the expressions for the finite distances
in Eqs. (\ref{BD7}) and (\ref{asslength}), respectively, that we have obtained
in the previous section. Second, in addition to the fact that
$d_{\mathrm{Euclid}}\left(  P_{i}\text{, }P_{j}\right)  \leq
d_{\mathrm{Taxicab}}\left(  P_{i}\text{, }P_{j}\right)  \leftrightarrow
d_{\mathrm{Bures}}\left(  P_{i}\text{, }P_{j}\right)  \leq d_{\mathrm{Sj\ddot
{o}qvist}}\left(  P_{i}\text{, }P_{j}\right)  $ for arbitrary points $P_{i}$
and $P_{j}$, it can be noted that a given probe point in the Sj\"{o}qvist
manifold appears to be locally surrounded by a greater number of points at the
same distance from the source. This, in turn, can be regarded as an indicator
of the presence of a higher degree of complexity during the change from an
initial point (source state) to a final point (target state). Therefore, this
set of points of discussion that we are offering here seem to land additional
support to the apparent emergence of a softer degree of complexity in Bures
manifolds when compared with Sj\"{o}qvist manifolds \cite{cafaroPRD22}. For
clarity, we remark that the proof of the inequality\textbf{ }%
$d_{\mathrm{Euclid}}\left(  P_{i}\text{, }P_{j}\right)  \leq
d_{\mathrm{Taxicab}}\left(  P_{i}\text{, }P_{j}\right)  $\textbf{ }can be
found in any standard topology book, including Ref. \cite{croom23}. The proof
of the inequality\textbf{ }$d_{\mathrm{Bures}}\left(  P_{i}\text{, }%
P_{j}\right)  \leq d_{\mathrm{Sj\ddot{o}qvist}}\left(  P_{i}\text{, }%
P_{j}\right)  $\textbf{)}, instead, follows from the analyses presented in
Refs. \cite{erik20,alsing23}. In particular, its origin can be traced back to
the fact that both quantum metrics originate from a specific minimization
procedure that, for the Bures metric, occurs in a larger space of unitary
matrices. For technical details on this minimization procedure, we refer to
Refs. \cite{erik20,alsing23}. For completeness, we also point out that once we
find a single violation of either the former (classical) or the latter
(quantum) inequalities, we can find several sets of points that would yield
the same violation. From a classical geometry standpoint, this is a
consequence of the fact that distances are invariant under isometries. In
particular, limiting our discussion to the case at hand, any planar isometry
mapping input points in $%
\mathbb{R}
^{2}$ to output points in $%
\mathbb{R}
^{2}$ is either a pure translation, a pure rotation about some center, or a
reflection followed by a translation (i.e., a glide reflection). Moreover, the
composition of two isometries is an isometry. From a quantum standpoint,
instead, an isometry is an inner-product preserving transformation that maps,
in general, between Hilbert spaces with different dimensions. In the
particular scenario in which input and output Hilbert spaces have the same
dimensions, the isometry is simply a unitary operation. For a general
discussion on the role of isometries in quantum information and computation,
we refer to Refs. \cite{molnar03,iten16}. Finally, for an illustrative
visualization of the Euclidean, Taxicab, Bures, and Sj\"{o}qvist geometries
that summarizes most of our discussion points, we refer to Fig. $3$.
Interestingly, inspired by the expression of the Bures fidelity \textrm{F}%
$_{\mathrm{Bures}}(\vec{a}$, $\vec{b})=\left[  1-\mathcal{L}_{\mathrm{Bures}%
}^{2}(\vec{a},\vec{b})/\left(  \mathcal{L}_{\mathrm{Bures}}^{2}\right)
_{\max}\right]  ^{2}$\textbf{ }with $\left(  \mathcal{L}_{\mathrm{Bures}%
}\right)  _{\max}=\sqrt{2}$, one may think of considering a sort of
Sj\"{o}qvist fidelity given by \textrm{F}$_{\mathrm{Sj\ddot{o}qvist}}(\vec{a}%
$, $\vec{b})=\left[  1-\mathcal{L}_{\mathrm{Sj\ddot{o}qvist}}^{2}(\vec{a}%
,\vec{b})/\left(  \mathcal{L}_{\mathrm{Sj\ddot{o}qvist}}^{2}\right)  _{\max
}\right]  ^{2}$ with $\left(  \mathcal{L}_{\mathrm{Sj\ddot{o}qvist}}\right)
_{\max}=\pi/2$. From these fidelities, define the ratio $\mathcal{R}(\vec{a}$,
$\vec{b})\overset{\text{def}}{=}$\textrm{F}$_{\mathrm{Sj\ddot{o}qvist}}%
(\vec{a}$, $\vec{b})/$\textrm{F}$_{\mathrm{Bures}}(\vec{a}$, $\vec{b})$ with
$\vec{a}$, $\vec{b}$ as in Fig\textbf{.} $3$ (for example)\textbf{. }Then, one
can check that the area of the two-dimensional parametric region with
parameters\textbf{ }$r$ and $\theta$ \textbf{\ }and specified by the
conditions $0\leq\mathcal{R}(\vec{a}$, $\vec{b})\leq1$, i.e. the region where
Bures fidelity is larger than the Sj\"{o}qvist fidelity, is greater than
$50\%$ of the total accessible two-dimensional parametric region with area
given by $\pi$ (i.e., the Lebesgue measure $\mu_{\mathrm{Lebesgue}}\left(
\left[  0,1\right]  _{r}\times\left[  0,\pi\right]  _{\theta}\right)  $ of the
interval $\left[  0,1\right]  _{r}\times\left[  0,\pi\right]  _{\theta}%
$).\textbf{ }Therefore, this type of approximate reasoning can be viewed as a
semi-quantitiative indication of the higher degree of distinguishability of
mixed quantum states by means of the Bures metric. Clearly, a deeper
comprehension of these facts would require an analysis extended to arbitrary
initial parametric configurations along with a more rigorously defined version
of\textbf{ }$\mathcal{R}(\vec{a}$, $\vec{b})$. Nevertheless, we believe that
interesting insights emerge from our approximate semi-quantitative discussion
proposed here. Summing up, our investigation suggests that the higher
sensitivity of the length of geodesic paths connecting a given pair of initial
and final mixed states of a quantum system in the Bures case is caused by the
lower density of accessible final states that are equidistant from a chosen
initial source state. This lower density, in turn, can be attributed to the
shorter length of geodesic paths in the Bures case. Finally, the shortness of
these paths is a consequence of the manner in which the quantumness (or,
alternatively, non-classicality) of mixed quantum states is geometrically
quantified with the Bures metric \cite{cafaro23EPJ} (i.e., the above-mentioned
way characterized by a minimization procedure in a larger space of unitary
matrices \cite{alsing23}). We are now ready for our conclusions.

\section{Final Remarks}

In this paper, building on our recent works in Refs.
\cite{cafaroPRD22,cafaro23EPJ,alsing23}, we presented more comprehensive
discussion on the differences between the Bures and the Sj\"{o}qvist metrics
inside a Bloch sphere. First, inspired by the works by Caves and Braunstein in
Refs. \cite{sam95A,sam95B}, we offered a formal comparative analysis between
the two metrics by critically discussing three alternative interpretations for
each metric. For the Bures metric, the three interpretations appear in Eqs.
(\ref{burespherical}), (\ref{corre0}), and (\ref{hubi}). For the Sj\"{o}qvist
metric, instead, the corresponding three interpretations emerge from Eqs.
(\ref{sspherical}), (\ref{corre1}), and (\ref{buri}), respectively. Second, we
illustrated (Fig. $1$) in an explicit fashion the different behaviors of the
geodesic paths (Eqs. (\ref{ca4}) and (\ref{rerikGOOD}) for the Bures and
Sj\"{o}qvist metrics cases, respectively) on each one of the two metric
manifolds. Third, we compared (Fig. $2$) the finite distances between an
initial and final mixed state when calculated with the two metrics (Eqs.
(\ref{BD7}) and (\ref{asslength}) for the Bures and Sj\"{o}qvist metrics
cases, respectively). Thanks to Eqs. (\ref{BD7}) and (\ref{asslength}) for
$\mathcal{L}_{\mathrm{Bures}}(\vec{a}$, $\vec{b})$ and $\mathcal{L}%
_{\mathrm{Sj\ddot{o}qvist}}(\vec{a}$, $\vec{b})$, respectively, we were able
to provide some intriguing discussion points ( along with a visual aid coming
from Fig. $3$) concerning some similarities between classical (Euclidean,
Taxicab) metrics in $%
\mathbb{R}
^{2}$ and quantum (Bures, Sj\"{o}qvist) metrics inside the Bloch sphere. In
particular, we argued that the fact that the Sj\"{o}qvist metric yields longer
finite distances, denser clouds of states that are equidistant from a fixed
source state and, finally, an unnatural violation of distance-based relative
ranking of pairs of points inside the Bloch sphere is at the origin of the
higher degree of complexity of the Sj\"{o}qvist manifold compared with the
Bures manifold as reported in Ref. \cite{cafaroPRD22}.

In the usual three-dimensional physical space, we ordinarily state that the
reason why it is difficult to distinguish two points is because they are close
together. In classical and quantum geometry, one tends to invert this line of
reasoning and claim that two points on a statistical manifold must be very
close together because it is hard to differentiate them \cite{caticha12}. In
particular, within the geometry of mixed quantum states, increasing distance
seems to correspond to more reliable distinguishability \cite{sam94}. From
Figs. $3\mathrm{(c)}$ and $3\mathrm{(d)}$, we note that for a given accessible
region \textrm{I}$_{r}\times\mathrm{I}_{\theta}\overset{\text{def}}{=}\left[
0,1\right]  \times\left[  0,\pi\right]  $ the lower density of level curves in
the Bures case is consistent with the observed softening of the complexity of
motion on Bures manifolds compared with Sj\"{o}qvist manifolds
\cite{cafaroPRD22}. Indeed, considering points at the same distance from the
source state as indistinguishable and viewing indistinguishability as an
obstruction to the evolution to new distinguishable states to be traversed
before arriving at a possible target state, a lower degree of the complexity
of motion would correspond to an accessible region made up of a greater number
of discernible states. Loosely speaking, Sj\"{o}qvist manifolds have some sort
of \textquotedblleft\emph{quantum labyrinth}\textquotedblright\ structure
greater than the one corresponding to Bures manifolds. Therefore, one can risk
to encounter longer paths of indistinguishability and, thus, can necessitate
to explore larger accessible regions before landing to the sought target state
\cite{cafaroPRD22,stefano11,ali17,ali21}.

In this paper, for simplicity and without loss of generality, we focused on
the discussion of single-qubit geodesic curves connecting pairs of points in
the $xz$-plane of the Bloch ball. However, to enhance the visual appeal of our
study, it could be worthwhile exploring the possibility of visualizing the
geodesic evolution of the three-dimensional (real) Bloch vectors in order to
gain clearer insights into the behavior of mixed quantum states. We leave the
consideration of this intriguing line of research to future scientific
efforts. In this work, we also focused on geometric aspects of two specific
metrics for mixed quantum states. For a general discussion on the relevant
criteria an arbitrary quantum distance must satisfy in order to be both
experimentally and theoretically meaningful, we refer to Refs.
\cite{nielsen05,mendonca08}. In particular, for a discussion on how to
experimentally determine the Bures and Sj\"{o}qvist distances by means of
interferometric procedures, we refer to Refs. \cite{bart19} and
\cite{erik20,silva21}, respectively. Moreover, we emphasize that our work here
does not consider the role of space-time geometry, as the quantum metrics we
discuss are purely Riemannian. However, given some formal similarities between
the quantum Bures and the classical closed Robertson-Walker spatial geometries
(Appendix B), it would be interesting to begin from this formal link and
elaborate on it so to help shedding some light on how to construct suitable
versions of quantum space-time geometries that can incorporate relativistic
physical effects within the framework of quantum physics
\cite{peres04A,peres04B,mann12,alsing12,alsing05}.

\smallskip

In summary, despite its limitations, we hope our work will motivate other
researchers and pave the way to additional investigations on the interplay
between quantum mechanics, geometry, and topological arguments. From our
standpoint, we have strong reasons to believe this work will undoubtedly
constitute a solid starting point for an extension of our recent work in Ref.
\cite{cafaro23} on qubit geodesics on the Bloch sphere from optimal-speed
Hamiltonian evolutions to qubit geodesics inside the Bloch sphere. For the
time being, we leave a more in-depth quantitative discussion on these
potential geometric extensions of our analytical findings, including
generalizations to mixed state geometry and quantum evolutions in
higher-dimensional Hilbert spaces, to forthcoming scientific investigations.

\begin{acknowledgments}
P.M.A. acknowledges support from the Air Force Office of Scientific Research
(AFOSR). C.C. thanks N. Andrzejkiewicz, B. Glindmyer, E. Liriano, and C. Neal
for helpful discussions on Euclidean and Taxicab geometries. C.C. is also
grateful to the United States Air Force Research Laboratory (AFRL) Summer
Faculty Fellowship Program for providing support for this work. Any opinions,
findings and conclusions or recommendations expressed in this material are
those of the author(s) and do not necessarily reflect the views of the Air
Force Research Laboratory (AFRL). The work of O.L. is partially financed by
the Ministry of Higher Education and Science of the Republic of Kazakhstan,
Grant: IRN AP19680128. Finally, O.L. is grateful to the Department of Physics
of the Al-Farabi University for hospitality during the period in which the
idea of this manuscript was discussed.
\end{acknowledgments}

\bigskip

\bigskip\pagebreak

\appendix

\section{Rotation of Bloch vectors}

In this Appendix, we discuss some technical details needed to understand the
reason why to compare finite distances between mixed quantum states in the
Bloch ball calculated with the Bures and Sj\"{o}qvist metrics, it is
sufficient to focus solely on points in the $xz$-plane. As mentioned in
Section III, this is essentially due to two facts. First, one needs to recall
that distances are preserved under rotations.\ Second, given two mixed states
$\rho_{1}=\rho\left(  \vec{p}_{1}\right)  \overset{\text{def}}{=}\left(
\mathrm{I}+\vec{p}_{1}\cdot\vec{\sigma}_{1}\right)  /2$ and $\rho_{2}%
=\rho\left(  \vec{p}_{2}\right)  \overset{\text{def}}{=}\left(  \mathrm{I}%
+\vec{p}_{2}\cdot\vec{\sigma}_{2}\right)  /2$, one needs to exploit the fact
that it is possible to construct a suitable composition of two $\mathrm{SO}%
(3$; $%
\mathbb{R}
)$ rotations acting on arbitrary Bloch vectors for mixed states, say $\vec
{p}_{1}$ and $\vec{p}_{2}$, such that the distances $\mathcal{L}\left(
\vec{p}_{1}\text{, }\vec{p}_{2}\right)  =\mathcal{L}\left(  \vec{p}%
_{1,new}\text{, }\vec{p}_{2,new}\right)  $ with $\vec{p}_{1,new}$ and $\vec
{p}_{2,new}$ belonging to the $xz$-plane. In this Appendix, we wish to
explicitly present these two $\mathrm{SO}(3$; $%
\mathbb{R}
)$ rotations needed to accomplish this task.

In general, recall that the unitary evolution of the state $\rho\left(
0\right)  $ under the unitary evolution operator $U\left(  t\right)  $ can be
described by $\rho\left(  t\right)  =U\left(  t\right)  \rho\left(  0\right)
U^{\dagger}\left(  t\right)  $. In terms of the Bloch vectors $\vec{a}\left(
0\right)  $ and $\vec{a}\left(  t\right)  $ with $\rho\left(  0\right)
\overset{\text{def}}{=}\left[  \mathrm{I}+\vec{a}\left(  0\right)  \cdot
\vec{\sigma}\right]  /2$ and $\rho\left(  t\right)  \overset{\text{def}}%
{=}\left[  \mathrm{I}+\vec{a}\left(  t\right)  \cdot\vec{\sigma}\right]  /2$,
respectively, we have that $\vec{a}\left(  0\right)  $ evolves to $\vec
{a}\left(  t\right)  $ by following the transformation law $\vec{a}\left(
t\right)  =\mathcal{R}_{\hat{n}}\left(  \alpha\right)  \vec{a}\left(
0\right)  $. The quantity $\mathcal{R}_{\hat{n}}\left(  \alpha\right)  $ is an
$\mathrm{SO}\left(  3\text{; }%
\mathbb{R}
\right)  $ rotation about the $\hat{n}$-axis by an angle $\alpha$. In general,
the temporal dependence can be encoded into both $\hat{n}$ and $\alpha$. The
relation between the $\mathrm{SU}\left(  2\text{; }%
\mathbb{C}
\right)  $ counterclockwise rotation by the angle $\alpha$ about the axis
$\hat{n}$, $U\left(  \alpha\text{, }\hat{n}\right)  =e^{-i\frac{\alpha}{2}%
\hat{n}\cdot\vec{\sigma}}$, and the $\mathrm{SO}\left(  3\text{; }%
\mathbb{R}
\right)  $ rotation $\mathcal{R}_{\hat{n}}\left(  \alpha\right)  $ is given by
\cite{sakurai85},%
\begin{equation}
\left[  \mathcal{R}_{\hat{n}}\left(  \alpha\right)  \right]  _{ij}%
\overset{\text{def}}{=}\frac{1}{2}\mathrm{tr}\left[  \sigma_{i}U\left(
\alpha\text{, }\hat{n}\right)  \sigma_{j}U^{\dagger}\left(  \alpha\text{,
}\hat{n}\right)  \right]  \text{,} \label{foto1}%
\end{equation}
with $1\leq i$, $j\leq3$. In particular, using the explicit expression of
$U\left(  \alpha\text{, }\hat{n}\right)  $ \cite{sakurai85}%
\begin{equation}
U\left(  \alpha\text{, }\hat{n}\right)  =e^{-i\frac{\alpha}{2}\hat{n}\cdot
\vec{\sigma}}=\left(
\begin{array}
[c]{cc}%
\cos\left(  \frac{\alpha}{2}\right)  -i\sin\left(  \frac{\alpha}{2}\right)
n_{z} & -i\sin\left(  \frac{\alpha}{2}\right)  \left(  n_{x}-in_{y}\right) \\
-i\sin\left(  \frac{\alpha}{2}\right)  \left(  n_{x}+in_{y}\right)  &
\cos\left(  \frac{\alpha}{2}\right)  +i\sin\left(  \frac{\alpha}{2}\right)
n_{z}%
\end{array}
\right)  \text{,} \label{foto2}%
\end{equation}
the $\left(  3\times3\right)  $-rotation matrix $\mathcal{R}_{\hat{n}}\left(
\alpha\right)  $ in Eq. (\ref{foto2}) becomes%
\begin{equation}
\mathcal{R}_{\hat{n}}\left(  \alpha\right)  =\left(
\begin{array}
[c]{ccc}%
\cos\left(  \alpha\right)  +n_{x}^{2}\left[  1-\cos\left(  \alpha\right)
\right]  & n_{x}n_{y}\left[  1-\cos\left(  \alpha\right)  \right]  -n_{z}%
\sin\left(  \alpha\right)  & n_{x}n_{z}\left[  1-\cos\left(  \alpha\right)
\right]  +n_{y}\sin\left(  \alpha\right) \\
n_{y}n_{x}\left[  1-\cos\left(  \alpha\right)  \right]  +n_{z}\sin\left(
\alpha\right)  & \cos\left(  \alpha\right)  +n_{y}^{2}\left[  1-\cos\left(
\alpha\right)  \right]  & n_{y}n_{z}\left[  1-\cos\left(  \alpha\right)
\right]  -n_{x}\sin\left(  \alpha\right) \\
n_{z}n_{x}\left[  1-\cos\left(  \alpha\right)  \right]  -n_{y}\sin\left(
\alpha\right)  & n_{z}n_{y}\left[  1-\cos\left(  \alpha\right)  \right]
+n_{x}\sin\left(  \alpha\right)  & \cos\left(  \alpha\right)  +n_{z}%
^{2}\left[  1-\cos\left(  \alpha\right)  \right]
\end{array}
\right)  \text{.} \label{ro}%
\end{equation}
To show that \textrm{dist}$\left(  \rho_{1}\text{, }\rho_{2}\right)
=$\textrm{dist}$\left(  \rho_{1,new}\text{, }\rho_{2,new}\right)
\leftrightarrow$\textrm{dist}$\left(  \vec{p}_{1}\text{, }\vec{p}_{2}\right)
=$\textrm{dist}$\left(  \vec{p}_{1,new}\text{, }\vec{p}_{2,new}\right)  $, we
need to make explicit the sequential action of the two $\mathrm{SO}\left(
3\text{; }%
\mathbb{R}
\right)  $ rotation matrices that need to act simultaneously on the pair
$\left(  \vec{p}_{1}\text{, }\vec{p}_{2}\right)  $. This will be described in
two steps. In the first step, we transition from the arbitrary pair $\left(
\vec{p}_{1}\text{, }\vec{p}_{2}\right)  $ with $\vec{p}_{1}\overset
{\text{def}}{=}\left(  r_{1}\sin\left(  \theta_{1}\right)  \cos\left(
\varphi_{1}\right)  \text{, }r_{1}\sin\left(  \theta_{1}\right)  \sin\left(
\varphi_{1}\right)  \text{, }r_{1}\cos\left(  \theta_{1}\right)  \right)  $
and $\vec{p}_{2}\overset{\text{def}}{=}\left(  r_{2}\sin\left(  \theta
_{2}\right)  \cos\left(  \varphi_{2}\right)  \text{, }r_{2}\sin\left(
\theta_{2}\right)  \sin\left(  \varphi_{2}\right)  \text{, }r_{2}\cos\left(
\theta_{2}\right)  \right)  $, respectively, to the new pair $\left(  \vec
{p}_{1}^{\prime}\text{, }\vec{p}_{2}^{\prime}\right)  $ given by%
\begin{equation}
\vec{p}_{1}^{\prime}=\mathcal{R}_{\frac{\vec{p}_{1}\times\hat{z}}{\left\Vert
\vec{p}_{1}\times\hat{z}\right\Vert }}\left(  \theta_{1}\right)  \vec{p}%
_{1}\text{, and }\vec{p}_{2}^{\prime}=\mathcal{R}_{\frac{\vec{p}_{1}\times
\hat{z}}{\left\Vert \vec{p}_{1}\times\hat{z}\right\Vert }}\left(  \theta
_{1}\right)  \vec{p}_{2}\text{,} \label{hold1}%
\end{equation}
respectively. In Eq. (\ref{hold1}), we have $\vec{p}_{1}^{\prime}=r_{1}\hat
{z}$, $\vec{p}_{2}^{\prime}=\left(  r_{2}\sin\left(  \theta_{2}^{\prime
}\right)  \cos\left(  \varphi_{2}^{\prime}\right)  \text{, }r_{2}\sin\left(
\theta_{2}^{\prime}\right)  \sin\left(  \varphi_{2}^{\prime}\right)  \text{,
}r_{2}\cos\left(  \theta_{2}^{\prime}\right)  \right)  $, and the
$\mathrm{SO}\left(  3\text{; }%
\mathbb{R}
\right)  $ rotation matrix $\mathcal{R}_{\frac{\vec{p}_{1}\times\hat{z}%
}{\left\Vert \vec{p}_{1}\times\hat{z}\right\Vert }}\left(  \theta_{1}\right)
$ given by%
\begin{equation}
\mathcal{R}_{\frac{\vec{p}_{1}\times\hat{z}}{\left\Vert \vec{p}_{1}\times
\hat{z}\right\Vert }}\left(  \theta_{1}\right)  \overset{\text{def}}{=}\left(
\begin{array}
[c]{ccc}%
\cos\left(  \theta_{1}\right)  +\sin^{2}(\varphi_{1})\left(  1-\cos\left(
\theta_{1}\right)  \right)  & -\sin(\varphi_{1})\cos(\varphi_{1})\left(
1-\cos\left(  \theta_{1}\right)  \right)  & -\cos(\varphi_{1})\sin\left(
\theta_{1}\right) \\
-\sin(\varphi_{1})\cos(\varphi_{1})\left(  1-\cos\left(  \theta_{1}\right)
\right)  & \cos\left(  \theta_{1}\right)  +\cos^{2}\left(  \varphi_{1}\right)
\left(  1-\cos\left(  \theta_{1}\right)  \right)  & -\sin(\varphi_{1}%
)\sin\left(  \theta_{1}\right) \\
\cos(\varphi_{1})\sin\left(  \theta_{1}\right)  & \sin(\varphi_{1})\sin\left(
\theta_{1}\right)  & \cos\left(  \theta_{1}\right)
\end{array}
\right)  \text{.}%
\end{equation}
In the second step, we transition from the pair $\left(  \vec{p}_{1}^{\prime
}\text{, }\vec{p}_{2}^{\prime}\right)  $ to the final pair $\left(  \vec
{p}_{1,new}\text{, }\vec{p}_{2,new}\right)  $ given by
\begin{equation}
\vec{p}_{1,new}=\mathcal{R}_{\hat{z}}\left(  \varphi_{2}^{\prime}\right)
\vec{p}_{1}^{\prime}\text{, and }\vec{p}_{2,new}=\mathcal{R}_{\hat{z}}\left(
\varphi_{2}^{\prime}\right)  \vec{p}_{2}^{\prime}\text{,} \label{hold2}%
\end{equation}
respectively. In Eq. (\ref{hold2}), we have $\vec{p}_{1,new}=\vec{p}%
_{1}^{\prime}=r_{1}\hat{z}$, $\vec{p}_{2,new}=\left(  r_{2}\sin\left(
\theta_{2}^{\prime}\right)  \text{, }0\text{, }\cos\left(  \theta_{2}^{\prime
}\right)  \right)  $, and the $\mathrm{SO}\left(  3\text{; }%
\mathbb{R}
\right)  $ rotation matrix $\mathcal{R}_{\hat{z}}\left(  \varphi_{2}^{\prime
}\right)  $ defined as%
\begin{equation}
\mathcal{R}_{\hat{z}}\left(  \varphi_{2}^{\prime}\right)  \overset{\text{def}%
}{=}\allowbreak\left(
\begin{array}
[c]{ccc}%
\cos\left(  \varphi_{2}^{\prime}\right)  & \sin\left(  \varphi_{2}^{\prime
}\right)  & 0\\
-\sin\left(  \varphi_{2}^{\prime}\right)  & \cos\left(  \varphi_{2}^{\prime
}\right)  & 0\\
0 & 0 & 1
\end{array}
\right)  \text{.}%
\end{equation}
Clearly, before ending our discussion, we need to specify how to obtain the
angles $\theta_{2}^{\prime}$ and $\varphi_{2}^{\prime}$. This can be easily
accomplished thanks to our knowledge of the Bloch vector $\vec{p}_{2}^{\prime
}=\left(  \vec{p}_{2}^{\prime}\cdot\hat{x}\text{, }\vec{p}_{2}^{\prime}%
\cdot\hat{y}\text{, }\vec{p}_{2}^{\prime}\cdot\hat{z}\right)  $. More
explicitly, the angles $\theta_{2}^{\prime}$ and $\varphi_{2}^{\prime}$ are
given by%
\begin{equation}
\theta_{2}^{\prime}\overset{\text{def}}{=}\arccos\left(  \frac{\vec{p}%
_{2}^{\prime}\cdot\hat{z}}{\left\Vert \vec{p}_{2}^{\prime}\right\Vert
}\right)  \text{, and }\varphi_{2}^{\prime}\overset{\text{def}}{=}\frac
{\vec{p}_{2}^{\prime}\cdot\hat{y}}{\left\vert \vec{p}_{2}^{\prime}\cdot\hat
{y}\right\vert }\arccos\left(  \frac{\vec{p}_{2}^{\prime}\cdot\hat{x}}%
{\sqrt{\left(  \vec{p}_{2}^{\prime}\cdot\hat{x}\right)  ^{2}+\left(  \vec
{p}_{2}^{\prime}\cdot\hat{y}\right)  ^{2}}}\right)  \text{,} \label{nakedhand}%
\end{equation}
respectively. Note that when $\vec{p}_{2}^{\prime}\cdot\hat{x}>0$,
$\varphi_{2}^{\prime}$ in Eq. (\ref{nakedhand}) reduces to $\arctan\left[
\left(  \vec{p}_{2}^{\prime}\cdot\hat{y}\right)  /\left(  \vec{p}_{2}^{\prime
}\cdot\hat{x}\right)  \right]  $ \cite{stewart21}. Thanks to the relations in
Eq. (\ref{nakedhand}), our discussion can be considered complete now.

\section{Bures geometry and closed Robertson-Walker spatial geometry}

In this Appendix, we discuss some similarities between the quantum Bures and
the classical closed Robertson-Walker spatial geometries. It is known that
geodesic paths encode relevant information about the curved space being
characterized by a proper metric. In general relativity, for instance,
geodesic paths extend the concept of straight lines to curved space-time. In
the geometry of quantum evolutions, instead, a geodesic is viewed as a path of
minimal statistical length that connects two quantum states along which the
maximal number of statistically distinguishable states is minimum. In
Einstein's general theory of relativity, the dynamical evolution of a physical
system is linked to the space-time geometry in a very neat
manner.\ Specifically, space-time explains matter how to move. Matter, in
turn, informs space-time how to curve. This link between matter and geometry
is neatly summarized in Einstein's field equations \cite{weinberg}, $\left[
\left(  8\pi G/c^{4}\right)  \mathcal{T}_{\mu\nu}\right]
_{\text{\textrm{matter}}}=\left[  \mathcal{R}_{\mu\nu}-(1/2)g_{\mu\nu
}\mathcal{R}\right]  _{\text{\textrm{geometry}}}$. In the previous relation,
$G$ is Newton's gravitational constant, $c$ is the speed of light in vacuum,
$\mathcal{T}_{\mu\nu}$ is the stress-energy tensor, $\mathcal{R}_{\mu\nu}$ is
the Ricci curvature tensor, $\mathcal{R}$ is the scalar curvature, and,
finally, $g_{\mu\nu}$ is the space-time metric tensor. In geometric
formulations of quantum mechanics, the geometry on the space of quantum
states, either pure \cite{wootters81} or mixed \cite{sam94}, specifies
limitations on our capacity of discriminating one state from another by means
of measurements. Therefore, unlike what happens in classical general
relativity, the geometry on the space of quantum states does no express, in
general, the actual dynamical evolution of a quantum system
\cite{sam95A,sam95B}.

Despite these differences, we remark that the Bures line element can be recast
as (neglecting the constant multiplicative factor $1/4$)%
\begin{equation}
ds_{\mathrm{Bures}}^{2}=d\chi^{2}+\sin^{2}\left(  \chi\right)  \left[
d\theta^{2}+\sin^{2}\left(  \theta\right)  d\varphi^{2}\right]  \text{,}
\label{corre00}%
\end{equation}
and is identical to the spatial metric component of the (closed) spherical
Robertson-Walker space-time metric that characterizes the so-called Freedman
model in cosmology. This space-time metric is given by \cite{weinberg},%
\begin{equation}
ds_{\mathrm{RW}}^{2}=-c^{2}dt^{2}+R^{2}\left(  t\right)  \left\{  \frac
{dr^{2}}{1-kr^{2}}+r^{2}\left[  d\theta^{2}+\sin^{2}\left(  \theta\right)
d\varphi^{2}\right]  \right\}  \text{,} \label{RW1}%
\end{equation}
where $R\left(  t\right)  $ is the cosmic scale factor and $k$\textbf{ }the
spatial curvature that can assume values $+1$ (spherical, or closed Universe),
$0$ (flat Universe), or $-1$ (hyperbolic, or open Universe). The dynamics of
the space-time geometry, once\textbf{ }$k$\textbf{ }is fixed, is fully
determined once the time-dependence of the cosmic scale factor is known
\cite{orlando12,orlando16}. Setting $t=t_{0}$, $R\left(  t_{0}\right)  =1$,
and $k=1$, Eq. (\ref{RW1}) reduced to the spatial metric%
\begin{equation}
dl_{\mathrm{RW}}^{2}=\frac{dr^{2}}{1-r^{2}}+r^{2}\left[  d\theta^{2}+\sin
^{2}\left(  \theta\right)  d\varphi^{2}\right]  \text{.} \label{RW2}%
\end{equation}
Then, performing a transformation from Cartesian to (spherical) polar
coordinates given by $x\overset{\text{def}}{=}\sin\left(  \chi\right)
\sin\left(  \theta\right)  \cos\left(  \varphi\right)  $, $y\overset
{\text{def}}{=}\sin\left(  \chi\right)  \sin\left(  \theta\right)  \sin\left(
\varphi\right)  $, $z\overset{\text{def}}{=}\sin\left(  \chi\right)
\cos\left(  \theta\right)  $, $w\overset{\text{def}}{=}\cos\left(
\theta\right)  $ \cite{MTW} (where, as usual, $\theta\in\left[  0,\pi\right]
$ is the polar angle, $\varphi\in\left[  0,2\pi\right)  $ is the azimuthal
angle, and $\chi\in\left(  0,\pi\right)  $ is the hyperspherical angle
\cite{ohanian}), the spatial line element $dl^{2}$ in Eq. (\ref{RW2}) becomes%
\begin{equation}
dl_{\mathrm{RW}}^{2}\overset{\text{def}}{=}d\chi^{2}+\sin^{2}\left(
\chi\right)  \left[  d\theta^{2}+\sin^{2}\left(  \theta\right)  d\varphi
^{2}\right]  \text{,} \label{RW3}%
\end{equation}
with $r\overset{\text{def}}{=}(x^{2}+y^{2}+z^{2})^{1/2}=\sin\left(
\chi\right)  $. The quantity $dl_{\mathrm{RW}}^{2}$ in Eq. (\ref{RW3})
coincides with $ds_{\mathrm{Bures}}^{2}$ in Eq. (\ref{corre00}) and represents
the metric of a three-sphere $S^{3}$ of unit radius. The three-sphere $S^{3}$
can be visualized as embedded in a four-dimensional Euclidean space and
specified by the condition $x^{2}+y^{2}+z^{2}+w^{2}=1$. Interestingly, just as
excursions off the three sphere $S^{3}$ are physically meaningless and
forbidden in general relativity \cite{MTW}, quantum systems in single-qubit
mixed states cannot escape from the inside of the Bloch sphere. Finally, for a
discussion on the measurement of lengths in curved space-time, we refer for
completeness to Refs. \cite{newman59,hans96,felice10,mac22}.


\begin{thebibliography}{99}                                                                                               %


\bibitem {karol06}I. Bengtsson and K. Zyczkowski, \emph{Geometry of Quantum
States}, Cambridge University Press (2006).

\bibitem {cafaroPRD22}C. Cafaro and P. M. Alsing, \emph{Complexity of pure and
mixed qubit geodesic paths on curved manifolds}, Phys. Rev. \textbf{D106},
096004 (2022).

\bibitem {chapman18}S. Chapman, M. P. Heller, H. Marrachio, and F. Pastawski,
\emph{Toward a definition of complexity for quantum field theory states},
Phys. Rev. Lett. \textbf{120}, 121602 (2018).

\bibitem {ruan21}S.-M. Ruan, \emph{Circuit Complexity of Mixed States}, Ph.D.
in Physics, University of Waterloo (2021).

\bibitem {gu12}M. Gu, K. Wiesner, E. Rieper, and V. Vedral, \emph{Quantum
mechanics can reduce the complexity of classical models}, Nature
Communications \textbf{3}, 762 (2012).

\bibitem {iaconis21}J. Iaconis, \emph{Quantum state complexity in
computationally tractable quantum circuits}, PRX Quantum \textbf{2}, 010329 (2021).

\bibitem {brandao21}F. G. S. L. Brandao, W. Chemissany, N. Hunter-Jones, R.
Kueng, and J. Preskill, \emph{Models of quantum complexity}, PRX Quantum
\textbf{2}, 030316 (2021).

\bibitem {bala22}V. Balasubramanian, P. Caputa, J. M. Magan, and Q. Wu,
\emph{Quantum chaos and the complexity of spread of states}, Phys. Rev.
\textbf{D106}, 046007 (2022).

\bibitem {belin22}A. Belin, R. C. Myers, S.-M. Ruan, G. Sarosi, and A. J.
Speranza, \emph{Does complexity equal anything?}, Phys. Rev. Lett.
\textbf{128}, 081602 (2022).

\bibitem {omidi23}F. Omidi, \emph{Generalized volume-complexity for two-sided
hyperscaling violating black branes}, J. High Energ. Phys. \textbf{01, }105 (2023).

\bibitem {karol98}K. Zyczkowski, P. Horodecki, A.\ Sanpera, and M. Lewenstein,
\emph{Volume of the set of separable states}, Phys. Rev. \textbf{A58}, 883 (1998).

\bibitem {karol99}K. Zyczkowski, \emph{Volume of the set of separable states}.
II, Phys, Rev. \textbf{A60}, 3496 (1999).

\bibitem {felice17}D. Felice, H. Q. Minh, and S. Mancini, \emph{The volume of
Gaussian states by information geometry}, J. Math. Phys.\textbf{\ 58}, 012201 (2017).

\bibitem {felice18}M. Rexiti, D. Felice, and S. Mancini, \emph{The volume of
two-qubit states by information geometry}, Entropy \textbf{20}, 146 (2018).

\bibitem {kz01}K. Zyczkowski and H.-J. Sommers, \emph{Induced measures in the
space of mixed quantum states}, J. Phys. A: Math. Gen. \textbf{34}, 7111 (2001).

\bibitem {kz03}H.-J. Sommers and K. Zyczkowski, \emph{Bures volume of the set
of mixed quantum states}, J. Phys. A: Math. Gen. \textbf{36}, 10083 (2003).

\bibitem {kz03A}K. Zyczkowski and H.-J. Sommers, \emph{Hilbert-Schmidt volume
of the set of mixed quantum states}, J. Phys. A: Math. Gen. \textbf{36}, 10115 (2003).

\bibitem {andai04}A. Andai, \emph{Volume of the quantum mechanical state
space}, J. Phys. A: Math. Gen. \textbf{39}, 13641 (2004).

\bibitem {ye09}D. Ye, \emph{On the Bures volume of separable quantum states},
J. Math. Phys. \textbf{50}, 083502 (2009).

\bibitem {ye10}D. Ye, \emph{On the comparison of volumes of quantum states},
J. Phys. A: Math. Theor. \textbf{43}, 315301 (2010).

\bibitem {singh14}R. Singh, R. Kunjwal, and R. Simon, \emph{Relative volume of
separable bipartite states}, Phys. Rev. \textbf{A89}, 022308 (2014).

\bibitem {siu20}K. Siudzinska, \emph{Geometry of generalized Pauli channels},
Phys. Rev. \textbf{A101}, 062323 (2020).

\bibitem {bures69}D. Bures, \emph{An extension of Kakutani's theorem on
infinite product measures to the tensor product of semifinite }$\omega^{\ast}%
$\emph{-algebras}, Trans. Amer. Math.\ Soc. \textbf{135}, 199 (1969).

\bibitem {uhlmann76}A. Uhlmann, \emph{The \textquotedblleft transition
probability\textquotedblright\ in the state space of a }$\ast$\emph{-algebra},
Rep. Math. Phys. \textbf{9}, 273 (1976).

\bibitem {hubner92}M. H\"{u}bner, \emph{Explicit computation of the Bures
distance for density matrices}, Phys. Lett. \textbf{A163}, 239 (1992).

\bibitem {sam94}S. L. Braunstein and C. M. Caves, \emph{Statistical distance
and the geometry of quantum states}, Phys. Rev. Lett. \textbf{72}, 3439 (1994).

\bibitem {erik20}E. Sj\"{o}qvist, \emph{Geometry along evolution of mixed
quantum states}, Phys. Rev. Research \textbf{2}, 013344 (2020).

\bibitem {cafaro23EPJ}C. Cafaro and P. M. Alsing, \emph{Bures and Sj\"{o}qvist
metrics over thermal state manifolds for spin qubits and superconducting flux
qubits}, Eur. Phys. J. Plus \textbf{138}, 655 (2023).

\bibitem {alsing23}P. M. Alsing, C. Cafaro, O. Luongo, C. Lupo, S. Mancini,
and H. Quevedo, \emph{Comparing metrics for mixed quantum states: Sj\"{o}qvist
and Bures}, Phys. Rev. \textbf{A107}, 052411 (2023).

\bibitem {cafaro07}C. Cafaro and S. A. Ali, \emph{Jacobi fields on statistical
manifolds of negative curvature}, Physica \textbf{D234}, 70 (2007).

\bibitem {cafarothesis}C. Cafaro, \emph{The Information Geometry of Chaos},
PhD Thesis, State University of New York at Albany, Albany-NY, USA (2008).

\bibitem {hubner93}M. H\"{u}bner, \emph{Computation of Uhlmann's parallel
transport for density matrices and the Bures metric on three-dimensional
Hilbert space}, Phys. Lett. \textbf{A179}, 226 (1993).

\bibitem {slater98}P. B. Slater, \emph{Bures metric for certain
high-dimensional quantum systems}, Phys. Lett. \textbf{A244}, 35 (1998).

\bibitem {dittmann99}J. Dittmann, \emph{Note on explicit formulae for Bures
metric}, J. Phys. \textbf{A32}, 2663 (1999).

\bibitem {zanardi07A}P. Zanardi, P. Giorda, and M. Cozzini,
\emph{Information-theoretic differential geometry of quantum phase
transitions}, Phys. Rev. Lett. \textbf{99}, 100603 (2007).

\bibitem {zanardi07B}P. Zanardi, L. Campos Venuti, and P. Giorda, \emph{Bures
metric over thermal manifolds and quantum criticality}, Phys. Rev.
\textbf{A76}, 062318 (2007).

\bibitem {safranek17}D. Safranek, \emph{Discontinuities of the quantum Fisher
information and the Bures metric}, Phys. Rev. \textbf{A95}, 052320 (2017).

\bibitem {hornedal22}N. Hornedal, D. Allan, and O. S\"{o}nnerborn,
\emph{Extensions of the Mandelstam-Tamm quantum speed limit to systems in
mixed states}, New J. Phys. \textbf{24}, 055004 (2022).

\bibitem {huang23}J.-H. Huang, S.-S. Dong, G.-L. Chen, N.-R. Zhou, F.-Y. Liu,
and L.-G. Qin,\emph{\ Path distance of a quantum unitary evolution}, Phys.
Rev. \textbf{A108}, 022204 (2023).

\bibitem {silva21}H. Silva, B. Mera, and N. Paunkovic, \emph{Interferometric
geometry from symmetry-broken Uhlmann gauge group with applications to
topological phase transitions}, Phys. Rev. \textbf{B103}, 085127 (2021).

\bibitem {dasilva21}H. V. da Silva, \emph{Quantum information geometry and
applications}, MS Thesis in Engineering Physics, IT Lisboa (2021).

\bibitem {kim21}E. Kim, \emph{Information geometry, fluctuations,
non-equilibrium thermodynamics, and geodesics in complex systems}, Entropy
\textbf{23}, 1393 (2021).

\bibitem {mera22}B. Mera, N. Paunkovic, S. T. Amin, and V. R. Vieira,
\emph{Information geometry of quantum critical submanifolds: Relevant,
marginal, and irrelevant operators}, Phys. Rev. \textbf{B106}, 155101 (2022).

\bibitem {daniel23}A. Daniel, C. Bruder, and M. Koppenh\"{o}fer,
\emph{Geometric phase in quantum synchronization}, Phys. Rev. Research
\textbf{5}, 023182 (2023).

\bibitem {chien23}X.-Y. Hou, Z. Zhou, X. Wang, H. Guo, and C.-C. Chien,
\emph{Local geometry and quantum geometric tensor of mixed states},
arXiv:quant-ph/2305.07597 (2023).

\bibitem {nielsen00}M. A. Nielsen and I. L. Chuang, \emph{Quantum Computation
and Quantum Information}, Cambridge University Press (2000).

\bibitem {sam95A}S. L. Braunstein and C. M. Caves, \emph{Geometry of quantum
states}, Annals of the New York Academy of Sciences \textbf{755}, 786 (1995).

\bibitem {sam95B}S. L. Braunstein and C. M. Caves, \emph{Geometry of quantum
states}. In: B. V. Belavkin, O. Hirota, and R. L. Hudson (Eds.), Quantum
Communications and Measurement, pp 21-30. Springer, Boston, MA (1995).

\bibitem {cafaro12}C. Cafaro and S. Mancini, \emph{Characterizing the
depolarizing quantum channel in terms of Riemannian geometry}, Int. J. Geom.
Meth. Mod. Phys. \textbf{9}, 1260020 (2012).

\bibitem {ole14a}O. Andersson and H. Heydari, \emph{Geometric uncertainty
relation for mixed quantum states}, J. Math. Phys. \textbf{55}, 042110 (2014).

\bibitem {ole14b}O. Andersson and H. Heydari, \emph{Quantum speed limits and
optimal Hamiltonians for driven systems in mixed states}, J. Phys. A: Math.
Theor. \textbf{47}, 215301 (2014).

\bibitem {ole15}O. Andersson and H. Heydari, \emph{A symmetry approach to
geometric phase for quantum ensembles}, J. Phys. A: Math. Theor. \textbf{48},
485302 (2015).

\bibitem {ole19}O. Andersson, \emph{Holonomy in Quantum Information Geometry},
Thesis for the degree of Licentiate of Philosophy in Theoretical Physics,
arXiv:quant-ph/1910.08140 (2019).

\bibitem {jozsa94}R. Jozsa, \emph{Fidelity for mixed quantum states}, Journal
of Modern Optics \textbf{41}, 2315 (1994).

\bibitem {wilde17}M. Wilde, \emph{Quantum Information Theory}, Cambridge
University Press (2017).

\bibitem {croom23}F. H. Croom, \emph{Principles of Topology}, Dover
Publications, Garden City, New York (2016).

\bibitem {molnar03}L. Molnar and W. Timmermann, \emph{Isometries of quantum
states}, J. Phys. A: Math. Gen. \textbf{36}, 267 (2003).

\bibitem {iten16}R. Iten, R. Colbeck, I. Kukuljan, J. Home, and M. Christandl,
\emph{Quantum circuit for isometries}, Phys. Rev.\ \textbf{A93}, 032318 (2016).

\bibitem {caticha12}A. Caticha, \emph{Entropic Inference and the Foundations
of Physics}, University of S\~{a}o Paulo Press, Brazil (2012).

\bibitem {stefano11}C. Cafaro and S. Mancini, \emph{Quantifying the complexity
of geodesic paths on curved statistical manifolds through information
geometric entropies and Jacobi fields}, Physica \textbf{D240}, 607 (2011).

\bibitem {ali17}S. A. Ali and C. Cafaro, \emph{Theoretical investigations of
an information geometric approach to complexity}, Rev. Math. Phys.
\textbf{29}, 1730002 (2017).

\bibitem {ali21}C. Cafaro and S. A. Ali, \emph{Information geometric measures
of complexity with applications to classical and quantum physical settings},
Foundations \textbf{1}, 45 (2021).

\bibitem {nielsen05}A. Gilchrist, N. K. Langford, and M. A. Nielsen,
\emph{Distance measures to compare real and ideal quantum processes}, Phys.
Rev. \textbf{A71}, 062310 (2005).

\bibitem {mendonca08}P. E. M. F. Mendonca, R. d. J. Napolitano, M. A.
Marchiolli, C. J. Foster, and Y.-C. Liang, \emph{Alternative fidelity measure
between quantum states}, Phys. Rev. \textbf{A78}, 052330 (2008).

\bibitem {bart19}K. Bartkiewicz, V. Travnicek, and K. Lemr, \emph{Measuring
distances in Hilbert space by many-particle interference}, Phys. Rev.
\textbf{A99}, 032336 (2019).

\bibitem {peres04A}A. Peres and D. R. Terno, \emph{Quantum information and
relativity theory}, Rev. Mod. Phys. \textbf{76}, 93 (2004).

\bibitem {peres04B}A. Peres, \emph{Quantum information and general
relativity}, Fortsch. Phys. \textbf{52}, 1052 (2004).

\bibitem {mann12}R. B.\ Mann and T. C. Ralph, \emph{Relativistic quantum
information}, Focus issue: Relativistic quantum information, Class. Quantum
Grav.\textbf{\ 29}, 220301 (2012).

\bibitem {alsing12}P. M. Alsing and I. Fuentes, \emph{Observer dependent
entanglement}, Focus issue: Relativistic quantum information, Class. Quantum
Grav. \textbf{29}, 224001 (2012).

\bibitem {alsing05}P. M.\ Alsing, J. P. Dowling, and G. J. Milburn, \emph{Ion
trap simulations of quantum fields in an expanding Universe}, Phys. Rev. Lett.
\textbf{94}, 220401 (2005).

\bibitem {cafaro23}C. Cafaro and P. M. Alsing, \emph{Qubit geodesics on the
Bloch sphere from optimal-speed Hamiltonian evolutions}, Class. Quantum Grav.
\textbf{40}, 115005 (2023).

\bibitem {sakurai85}J. J. Sakurai, \emph{Modern Quantum Mechanics}, Addison
Wesley Publishing Company, Inc. (1985).

\bibitem {stewart21}J. Stewart, D. Clegg, and S. Watson, \emph{Calculus: Early
Transcendentals}, Cengage Learning, Inc. (2021).

\bibitem {weinberg}S. Weinberg, \emph{Gravitation and Cosmology: Principles
and Applications of the General Theory of Relativity}, John Wiley and Sons,
Inc. (1972).

\bibitem {wootters81}W. K. Wootters, \emph{Statistical distance and Hilbert
space}, Phys. Rev. \textbf{D23}, 357 (1981).

\bibitem {orlando12}A. Aviles, C. Gruber, O. Luongo, and H. Quevedo,\emph{
Cosmography and constraints on the equation of state of the Universe in
various parametrizations}, Phys. Rev. \textbf{D86}, 123516 (2012).

\bibitem {orlando16}P. K. S. Dunsby and O. Luongo, \emph{On the theory and
applications of modern cosmography}, Int. J. Geom. Meth. Mod. Phys.
\textbf{13}, 1630002 (2016).

\bibitem {MTW}C. W. Misner, K. S. Thorne, and J. A. Wheeler,
\emph{Gravitation}, W. H. Freeman \& Company (1973).

\bibitem {ohanian}H. O. Ohanian and R. Ruffini, \emph{Gravitation and
Spacetime}, W. W. Norton \& Company (1976).

\bibitem {newman59}E. Newman and J. N. Goldberg, \emph{Measurement of distance
in general relativity}, Phys. Rev. \textbf{114}, 1391 (1959).

\bibitem {hans96}H.-J. Schmidt, \emph{How should we measure spatial
distances?}, Gen. Rel. Grav.\textbf{\ 28}, 899 (1996).

\bibitem {felice10}F. De Felice and D. Bini,\emph{\ Classical measurements in
curved spacetime}, Cambridge University Press (2010).

\bibitem {mac22}C. MacLaurin, \emph{Clarifying spatial distance measurement},
Proceedings of the Fifteenth Marcel Grossman Meeting, pp. 1372-1377 (2022).
\end{thebibliography}
\end{document}